# Networks and the Resilience and Fall of Empires:
# a Macro-Comparison of the Imperium Romanum and Imperial China

Johannes Preiser-Kapeller

Email: Johannes.Preiser-Kapeller@oeaw.ac.at

**Abstract:** This paper proposes to proceed from a rather metaphorical application of network terminology on polities and imperial formations of the past to an actual use of tools and concepts of network science. For this purpose, a well-established network model of the route system in the Roman Empire (ORBIS) and a newly created network model of the infrastructural web of Imperial China are visualised and analysed with regard to their structural properties. Findings indicate that these systems could be understood as large-scale complex networks with pronounced differences in centrality and connectivity among places and a hierarchical sequence of clusters across spatial scales from the over-regional to the local level. Such properties in turn would influence the cohesion and robustness of imperial networks, as is demonstrated with two tests on the model´s vulnerability to node failure and to the collapse of long-distance connectivity. Tentatively, results can be connected with actual historical dynamics and thus hint at underlying network mechanisms of large-scale integration and dis-integration of political formations.

**Introduction**

In her 2009 book on "Law and Geography in European Empires", *Lauren Benton* (2009, p. 2) stated: "*Empires did not cover space evenly but composed a fabric that was full of holes, stitched together out of pieces, a tangle of strings. Even in the most paradigmatic cases, an empire's spaces were politically fragmented; legally differentiated; and encased in irregular, porous, and sometimes undefined borders. Although empires did lay claim to vast stretches of territory, the nature of such claims was tempered by control that was exercised mainly over narrow bands, or corridors, and over enclaves and irregular zones around them.*"

This observation on more "cobwebby" spatial manifestations of imperial rule can be connected with earlier studies of *Monica L. Smith* (2005 and 2007), who borrowed concepts from ecology in order to characterize ancient states and empires as a "*series of nodes (population centres and resources) joined through corridors (roads, canals, rivers)*". Along similar lines (but without reference to Smith), Pekka Hämäläinen wrote that what he calls the "Comanche Empire" in the 18th-19th century North American West "*rested not on sweeping territorial control but on a* *capacity to connect vital economic and ecological nodes—trade corridors, grassy river valleys, grain-producing peasant villages, tribute-paying colonial capitals*" (*Hämäläinen* 2013; *St. John* 2013). Moreover, even for modern-day polities and international systems, *Parag Khanna* (2016) has argued for replacing traditional cartographic approaches with a "*connectography*" of routes, webs and corridors.

When referring to "nodes" and "connections", these authors (deliberately or unintentionally) use terminology from network theory; some, such as *Smith* (2005 and 2007) and others (*Glatz* 2009), have also argued for perceiving and depicting empires as networks. Most of them, however, resort to verbal descriptions or graphic visualisations, at best, without applying tools of actual network analysis, although there exists a number of such studies, dating back to the 1960s (*Carter* 1969; *Gorenflo and Bell* 1991).

On the following pages, we will demonstrate the potential to understand and to model large-scale polities of the past as networks with a comparison of two "paradigmatic" cases of imperial formations, the Roman Empire and Imperial China. Various recent volume have been devoted to a comparison of various aspects of these two empires, which even had (infrequent and tentative) contacts which each other (*Scheidel* 2009; *Mutschler and Mittag* 2008; *Auyang* 2015). Network theory, however, provides a different and common analytical basis beyond disciplinary boundaries for comparison. As will also become evident on the following pages, Rome and China of course very much differed in the "logics" of imperial connectivity, with the Imperium Romanum centred on a maritime Mediterranean core, while the sea marked more of a border for China with its web of terrestrial and riverine routes (*Mote* 1999; *Tuan* 2008; *Marks* 2017). Yet, both imperial formations under pre-modern technological conditions integrated enormous territories (of ca. five million km² each, see *Ruffing* 2012, p. 32; *Auyang* 2015, pp. 4-5) and had a lasting effect on further developments in Western and Eastern Afro-Eurasia (*Preiser-Kapeller* 2018).

**Some basic concepts and tools of network analysis**

Network theory assumes "*not only that ties matter, but that they are organised in a significant way, that this or that (node) has an interesting position in terms of its ties.*"(*Lemercier* 2012, p. 22) One central aim of network analysis is the identification of structures of relations, which emerge from the sum of interactions and connections between individuals, groups or sites and at the same time influence the scope of actions of everything and everyone entangled in such relations. For this purpose, data on the categories, intensity, frequency and dynamics of interactions and relations between entities of interest is systematically collected, allowing for further mathematical analysis. This information is organised in the form of matrices (with rows and columns) and graphs (with nodes [representing the elements to be connected] and edges [or links, representing the connections of interest]). Matrices and graphs are not only instruments of data collection and visualisation, but also the basis of further mathematical operations (*Wassermann and Faust* 1994, pp. 92-166; *Prell* 2012, pp. 9-16; *Barabási* 2016, pp. 42-67; for applications in archaeology and history: *Brughmans* 2012; *Knappett* 2013; *Collar, Coward, Brughmans and Mill* 2015; *Brughmans, Collar and Coward* 2016).

A quantifiable network model thus created allows for a structural analysis on three main levels (*Collar, Coward, Brughmans and Mill* 2015):

\* the level of single nodes. Respective measures take into account the immediate "neighbourhood" of a node – such as "degree", which measures the number of direct links of a node to other nodes (*Wassermann and Faust* 1994, pp. 178-183; *de Nooy, Mrvar and Batagelj* 2005, pp. 63-65; *Newman* 2010, pp. 168-169; *Prell* 2012, pp. 96-99). "Betweenness" measure the relative centrality of a node within the entire network due to its position on many or few possible paths between nodes otherwise unconnected. We interpret it as a potential for intermediation, while nodes with a high betweenness also provide cohesion and connectivity within the network (*Wassermann and Faust* 1994, pp. 188-192; *de Nooy, Mrvar and Batagelj* 2005, pp. 131-133; *Newman* 2010, pp. 185-193; *Prell* 2012, pp. 103-107). A further indicator of centrality is "closeness", which measures the length of all paths between a node and all other nodes. The "closer" a node is the lower is its total and average distance to all other nodes. Closeness can also be used as a measure of how fast it would take to spread resources or information from a node to all other nodes or how easily a node can be reached (and supplied with signals or material flows) from other nodes (*Wassermann and Faust* 1994, pp. 184-188; Prell 2012, pp. 107-109).

\* the level of groups of nodes, especially the identification of "clusters", meaning the existence of groups of nodes more densely connected among each other than to the rest of the network. A measure of the degree to which nodes in a graph tend to cluster together is the "clustering coefficient" (with values between 0 and 1) (*Wassermann and Faust* 1994, pp. 254-257). In order to detect such clusters, an inspection of a visualisation of a network can be already quite helpful; common visualisation tools arrange nodes more closely connected near to each other ("spring embedder"-algorithms) and thus provide a good impression of such structures (*Krempel* 2005; *Dorling* 2012). For exact identification, there exist various algorithms of "group detection", which aim at an optimal "partition" of the network (*de Nooy, Mrvar and*

*Batagelj* 2005, pp. 66-77; *Newman* 2010, pp. 372-382; *Prell* 2012, pp. 151-161; *Kadushin* 2012, pp. 46-49).

\* The level of the entire network: basic key figures are the number of nodes and of links, the maximum distance between two nodes (expressed in the number of links necessary to find a path from one to the other; "diameter") and the average distance (or path length) between two nodes. A low average path length among nodes together with a high clustering coefficient can be connected to the model of a "small world network", in which most nodes are linked to each other via a relatively small number of edges (*de Nooy, Mrvar and Batagelj* 2005, pp. 125-131; *Prell* 2012, pp. 171-172; *Watts* 1999). "Density" indicates the ratio of possible links actually present in a network: theoretically, all nodes in a network could be connected to each other (this would be a density of "1"). A density of "0.1" indicates that 10 % of these possible links exist within a network. The higher the number of nodes, the higher of course the number of possible links. Thus, in general, density tends to decrease with the size of a network. Therefore, it only makes sense to compare the densities of networks of (almost) the same size. Density can be interpreted as one indicator for the relative "cohesion", but also for the "complexity" of a network (*Prell* 2012, pp. 166-168; *Kadushin* 2012, p. 29). Other measurements are based on the equal or unequal distribution of quantitative characteristics such as degree, betweenness or closeness among nodes; a high "degree centralisation" indicates that many links are concentrated on a relatively small number of nodes, for instance (Prell 2012, pp. 168-170). These distributions can also be statistically analysed and visualised for all nodes (by counting the frequency of single degree values) and used for the comparison of networks (see **fig. 1** and **fig. 2**). Certain highly unequal degree distribution patterns (most prominently, power laws) have been interpreted as "signatures of complexity" of a network (Newman 2010, pp. 243-261).

The modelling of networks of routes between places demands further specifications. Links in such model are both weighted (meaning that a quantity is attributed to them) and directed (a link leads from point A to point B, for instance). The aim is to integrate aspects of what *Leif Isaksen* (2008) has called "transport friction" into calculations; otherwise, the actual costs of communication and exchange between sites, which influenced the frequency and strength of connections, would be ignored in network building. Links could be weighted by using the inverted geographical distance between them, for instance; thus, a link would be the stronger the shorter the distance between two nodes ("distant decay" effect). However, of course, if possible either existing information on the (temporal or economical) costs for using specific routes could be used. Otherwise, cost calculation stemming from GIS-based modelling of terrain and routes can be integrated. In riverine transport networks directed links leading upstream (from point A to point B) would be weighted differently from links leading downstream (from point B to point A) (*Rodrigue, Comtoi and Slack* 2013, pp. 307-317; *Taafee and Gauthier* 1973, pp. 100-158; *Ducruet and Zaidi* 2012; *Barthélemy* 2011; for historical transport networks cf. *Carter* 1969; *Pitts* 1978; *Gorenflo and Bell* 1991; *Graßhoff and Mittenhuber* 2009; *Leidwanger, Knappett et al.* 2014; *van Lanen et al.* 2015). Furthermore, there also exist measures especially developed for transport networks such as circuitry (or "alpha-index"). It measures the share of the maximum number of cycles or circuits (= a finite, closed path in which the initial node of the linkage sequence coincides with the terminal node) actually present in a traffic network model and thus indicates the existence of additional or alternative paths between nodes in the network and its relative connectivity and complexity (*Rodrigue, Comtoi and Slack* 2013, pp. 310-315; *Taaffe and Gauthier* 1973, pp. 104-105; *Wang, Ducruet and Wang* 2015, p. 455).

**Complex network models of empires**

As mentioned above, one of the earliest studies in the field of historical network research focused on the analysis of an imperial formation. In 1969, *F.W. Carter* created a network model of the route system in the Serbian Empire of Stefan Uroš IV Dušan (r. 1331-1355 CE), using the most important urban centres as nodes and the main trade routes as links. This paper also took into account the actual geographical distances between places. Since then, as outlined above, various approaches to model "transport friction" have been implemented in studies of historical route systems, ranging from the local to the regional and even "imperial" scale.

The most exhaustive network model of historical sea- and land routes of the Imperium Romanum is the "ORBIS Stanford Geospatial Network Model of the Roman World", developed by *Walter Scheidel* and *Elijah Meeks* (2014) in order to estimate transport cost and spatial integration within the Roman Empire. ORBIS is based on a network of roads, river and sea routes (in total, 1104 links) between 678 nodes (places), weighted according to the costs of transport (see **table 1** and **fig. 3**).[1] Since it is aiming at the entirety of the empire´s traffic system, ORBIS is less detailed on the regional and local level than network models for smaller areas (see for instance *Orengo and Livarda* 2016). We have corrected this data (especially with regard to the localisations of some places) and modified the network model, so that the link between two nodes (places) is the stronger the smaller the costs of overcoming the distance between them is, thus reflecting the ease or difficulty of transport and mobility between two localities (see also *Preiser-Kapeller* 2019).

---

[1] The data set was downloaded from: https://purl.stanford.edu/mn425tz9757 (Creative Commons Attribution 3.0 Unported License). For a similar model, see also *Graham* 2006. We are preparing the data for the Chinese network model to be made available for free via https://github.com.

For imperial China, unfortunately there does not exist any similar geospatial network model. *Chengjin Wang*, *César Ducruet* and *Wang Wei* in 2015 tried to reconstruct the road networks in China from 1600 BCE to 1900 CE on the basis of historical data and maps, but (to our knowledge) did not make their data available as Scheidel and Meek did. The same is true for an earlier study of *Wang and Ducruet* (2013) on the Chinese port system between 221 BCE and 2010 CE. A rich set of historical-geographical data across periods, however, is provided online via the China Historical Geographic Information System (CHGIS) hosted at Harvard University.[2] Based on this data, we constructed a network model of the most important riverine and land routes covering the historical central provinces of imperial China (cf. also *Brook* 1998). The model includes 1034 nodes (places) and 3034 links, weighted according to the cost of transport (based on geographical distance) (see **table 2** and **fig. 4**). Similar to the ORBIS-model, it aims to cover the entirety of the pre-modern traffic system in the Chinese Empire and is thus less detailed on the regional and local level. Equally, in the absence of suitable data, maritime links are missing; due to their relatively peripheral role (when compared with the Roman Mediterranean, for instance), we assume that still essential aspects of the transport network can be captured with our model (*Brook* 1998, pp. 615-619; *Wang and Ducruet* 2013; *Schottenhammer* 2015; see also the pioneering studies of *Skinner* 1977).

Networks are of course dynamic: relationships may be established, maintained, modified or terminated; nodes appear in a network and disappear (also from the sources). The common solution to capture at least part of these dynamics is to define "time-slices" (divided through meaningful caesurae in the development of the object of research, as defined by the researcher knowing the material) and to model distinct networks for each of them. Yet, since we reckon with a relatively long-term stability of core elements of the route and infrastructure networks we try to model (and also for the sake of simplicity), we decided to use static models (*de Nooy, Mrvar and Batagelj* 2005, pp. 92-95; *Lemercier* 2012, pp. 28-29; *Batagelj et al.* 2014). Equally, routes and infrastructures are only one "layer" of the various networks spanning across an imperial space, such as administration, commerce or religion. All these categories of connections could be integrated into a "multi-layer" network model, but unfortunately, we do not possess the same density of evidence for them across the entire empires as we have for the routes. At the same time, these flows of people and ideas were much more volatile than the infrastructural web, on which all of these other categories of linkages in turn were depending. As we will show below for the case of epidemics, the structure of the underlying route networks also influenced the pattern of diffusion of these other imperial network "layers" (*Collar* 2013; *Auyang* 2015; on multi-layer networks see *Bianconi* 2018). Against this background, we stress the famous aphorism that "all models are wrong but some are (hopefully) useful".[3]

---

[2] http://sites.fas.harvard.edu/~chgis/.

In a first step, we analysed the spatial and statistical distribution of measures of centrality among the nodes (= places) of the network. The number and accumulated strength of links of a node (its weighted degree, cf. *Newman* 2010, pp. 168-169) are high, where many localities are connected among each other over short distances via the most convenient transport medium – in the Roman case the sea (such as in the Aegean), in China riverine routes (such as in the densely settled delta area of the Yángzǐ Jiāng) (see **fig. 5** and **fig. 6**). Statistically, the distribution of these degree values is very unequal, with a high number of nodes with relatively low degree centrality and a small number of "hubs" with high centrality values (see **fig. 1** and **fig. 2**). As indicated above, betweenness measures the relative centrality of a node in the entire network due to its position on many (or few) potential shortest paths between nodes (Newman 2010, pp. 185-193). In the ORBIS-network, hubs of maritime transport (such as Alexandria, Rhodes or Messina) serve as most important integrators of the entire system in this regard. Other hotspots of connectivity are located at the main routes between the Northern provinces and the Mediterranean core (between Gaul and Spain, Central Europe and Italy, or the Lower Danube and the Aegean respectively Constantinople) (see **fig. 7**). In the China-network, the intermediary zones between West and East along the main riverine arteries of the Huang He and the Yángzǐ Jiāng can be identified as betweenness hotspots (see **fig. 8**). At the same time, the statistical distribution of betweenness values is even more unequal than the one of degree centrality (see **fig. 1** and **fig. 2**). "Closeness" measures the average length of all paths between a node and all other nodes in a network and indicates its overall "reachability" (or remoteness) (*Wassermann and Faust* 1994, pp. 184-188). Statistically, closeness-values are relatively equally distributed (see **fig. 1** and **fig. 2**). Their spatial distribution in the ORBIS-model indicates again the significant role of both maritime connectivity as well as of the intermediary regions between the Northern provinces and the Mediterranean core for the cohesion of the network (see **fig. 9**). In the China network, again the regions along the main riverine routes as well as the Grand Canal (constructed to connect the two stream systems of the Huang He and the Yángzǐ Jiāng especially for feeding the imperial capitals since the 7$^{th}$ century CE, see below) are privileged regarding closeness centrality (see **fig. 10**). Both networks are equally characterised by (relatively to their size) high values of circuitry (see **table 1** and **2**) (for similar findings cf. *Wang, Ducruet and Wang* 2015).

---

[3] The full version of this saying attributed to the statistician George Box sounds even more flattering for our purpose: "Since all models are wrong the scientist cannot obtain a "correct" one by excessive elaboration. On the contrary, following William of Occam he should seek an economical description of natural phenomena. Just as the ability to devise simple but evocative models is the signature of the great scientist so over-elaboration and over parameterization is often the mark of mediocrity." Cf. https://en.wikipedia.org/wiki/All_models_are_wrong.

The spatial distributions of network centrality measures in the two models confirm well-established assumptions on the inherent transport logics of these imperial systems. Their statistical distribution patterns (see **fig. 1** and **fig. 2**) can equally be interpreted as "signatures of complexity", thus identifying both imperial systems (or their imperfect reproduction in the two models) as "large scale complex networks" (for similar distribution patterns emerging from an analysis of Roman urbanisation in Asia Minor cf. *Hanson* 2011). They show "non-trivial" topological properties and patterns of connectivity between their nodes "that are neither purely regular nor purely random"; besides the high inequality in the distributions of centrality measures, these include a high value of circuitry and (as we will demonstrate below) a hierarchical community structure. These properties also allow for some assumptions on the overall robustness and tolerance towards the failure of nodes or links of these networks (*Newman* 2010, pp. 591-625; *Estrada* 2012, pp. 187-214; *Barabási* 2016, pp. 113-145, 271-305 and 321-362; *Preiser-Kapeller* 2015).

**Empire-wide connectivity, imperial capitals and ecologies**

A main aim of the above-mentioned pioneering study of *F. W. Carter* (1969, pp. 54-55) was to "learn more about the position" of the "successive capitals" within the route network of the Serbian Empire and "whether Stefan Dušan made the right choice in Skopje as his capital". Tsar Dušan did not, according to the findings of Carter, and thus (in Carter´s opinion) diminished the prospect for a sustained existence of his empire, since his residence of choice did not rank among the most central nodes in the network model. Other places would have been better situated, Carter argued, and thus would have provided better opportunities for economic development, the ease of "troop movement" as well as for the flows of materials.

Especially the aspect of material flows can be connected with the more recent concept of "imperial ecology", which *Sam White* (2011) in his study on the Ottoman Empire in the 16$^{th}$ and 17$^{th}$ century CE has defined as the "*particular flows of resources and population directed by the imperial center*" on which its success and survival depended. Within the web of the imperial ecology, in turn, the supply of the imperial centre (what has been called its "urban metabolism") can be identified as a core element (*González de Molina and Toledo* 2014; *Forman* 2014; *Schott* 2014). With regard to its dependency on the scale and reach of its network, *Peter Baccini* and *Paul H. Brunner* (2012, p. 58) made clear that the city of Rome in the imperial period had become "*an example of a system that could only maintain its size (…) on the basis of a political system that guaranteed the supply flows*" (see also *Morley* 1996; *Fletcher* 1995). The administrators and later the emperors of Rome invested heavily into the infrastructure of the city. Roman roads were built especially for military purposes (beginning with the Via Appia in 312 BCE leading from Rome to Capua and in 190 BCE expanded towards Brundisium at the Adriatic Sea); its maximum extent was 80,000 to 100,000 km (*Kolb* 2000; *Sauer* 2006; *Schneider* 2007, pp. 72-75, 89; *Ruffing* 2012, pp. 42-43; *Klee* 2010). Maritime links on the other hand served for the transport of bulk goods and became increasingly important for the provision of the growing capital. Since 123 BCE Rome became dependent on consignments of grain from North Africa, which at that time were financed with the taxes from the recently acquired territories in Western Asia Minor, thus establishing an early core triangle of flows of the imperial ecology (*Erdkamp* 2005; *Ruffing* 2012, pp. 98-99; *Sommer* 2013, pp. 90-91). Rome´s earlier harbour of Ostia was augmented with the immense artificial installations at Portus under Emperor Claudius (41-54 CE), later expanded by Emperor Trajan (98-117 CE) (*Keay* 2012; *Davies* 2005; *Scheidel, Morris and Saller* 2007, pp. 570-618). Thus, it was pivotal both for the coordination of imperial rule as well as for the very existence of the city that Rome remained well positioned in the growing network of routes. The orientation of these networks onto the capital was also symbolised with the *Milliarium Aureum*, the golden milestone erected by Emperor Augustus in 20 BCE on the Forum Romanum as point of origin of all roads in the empire (*Kolb* 2000; *Klee* 2010; *Temin* 2013). In addition, an analysis of the centrality measures for Rome within the ORBIS-network demonstrates its high connectivity, both with regard to betweenness as well as closeness values (see **table 3**; cf. also *Morley* 1996, pp. 63-68). For the entire network at large, however, ranging from Britannia to Egypt and from Spain to the Euphrates, Rome is not necessarily the most central hub anymore (see **table 3**). Especially after the severe military and political crisis of the 3[rd] century CE, strategic considerations demanded the relocation of imperial residences also on a more permanent basis to places near the endangered frontier zones (*Johne* 2008; *Pfeilschifter* 2014). These cities such as Milan, Aquileia, Sirmium or Serdica also show up in the analysis of the network model as well connected with regard to their betweenness and closeness values (see **table 3**), situated in the above-mentioned intermediary areas between the northern frontier and the Mediterranean core (see **fig. 7** and **fig. 9**). In terms of urban scale and population size, these places of course could not compete with Rome, which remained privileged with regard to the inflows of supplies from across the Mediterranean (*Erdkamp* 2005; *Scheidel, Morris and Saller* 2007, pp. 651-671). It was only the "new Rome" of Constantinople, inaugurated by Emperor Constantine I in 330 CE, which eventually would outperform the old capital at the Tiber also in these aspects in the 5$^{th}$ century CE. Constantinople is also the only one among the eleven imperial capitals in our sample, which ranks in the ORBIS-network model among the top ten in all three centrality measures of degree, betweenness and closeness (see **table 3**). This may contribute to an explanation of its long-time "success" as imperial centre over almost 1600 years until 1923 CE (the fall of the Ottoman dynasty), much longer than Rome itself (*Teall* 1959; *Mango and Dagron* 1995; *Preiser-Kapeller* 2018, pp. 249-250).

Similar observations can be made for China; the cities of Chang´an and Luoyang (see **fig. 4**) had served as capitals in the period of the Han-dynasty (206 BCE to 220 CE) and as starting points for the emerging imperial road system (which under the Han extended over 35,000 km). Both were located at rivers, which would allow for an easier inflow of resource into these urban centres that competed in scale with imperial Rome (*Lewis* 2007; *Tuan* 2008, pp. 75-81; *Wang, Ducruet and Wang* 2015, pp. 457-465; *Auyang* 2015, pp. 152-154; *Marks* 2017, pp. 90-93). Moreover, both Chang´an and Luoyang were selected as residences again after the "re-unification" of China by the Sui in the late 6$^{th}$ and early 7$^{th}$ century CE. Emperor Yang (604-617 CE) in an edict of 17 December 604 CE made clear that his decision to establish his residence in Luoyang was based on both tradition and its position in the network of imperial ecology: „*Luoyang has been a capital since antiquity. Within the precincts of its royal territory, Heaven and Earth merge with each other; yin and yang work in harmony. Commanding the Sanhe region, it is safeguarded by four mountain passes. With excellent land and water transportation, it provides a whole gamut of taxes and tribute*" (*Xiong* 2006, pp. 76-78). Yang and his father Emperor Wendi (581-604 CE) also began the construction of the Grand Canal system (see **fig. 4**), which connected the new breadbaskets in the South along the Yángzǐ Jiāng with the traditional imperial centres in the North and became a main lifeline of the imperial ecology for the next Millennium. The stress on society created by all these large-scale building projects together with a series of costly and unsuccessful military campaigns eventually contributed to the fall of Emperor Yang. Yet, the succeeding Tang Emperors equally built on the same system of capitals and metabolic flows (*Elvin* 1973, pp. 131-145; *Tuan* 2008, pp. 94-100; *Xiong* 2006; *Xiong* 2017; *Lewis* 2009, pp. 86-101 and 113-118; *Wang and Ducruet* 2013; *Marks* 2017, pp. 135-137). In terms of network analysis, both Chang´an and Luoyang are situated in a corridor of high betweenness at the main West-East axis of the Northern Chinese heartlands along the Huang He (see **fig. 8** and **fig. 10**). Luoyang, however, ranks higher than Chang´an with regard to closeness centrality and far higher in its degree-value, which reflects its more direct integration into the Grand Canal system (see **table 4**). Luoyang is equally identified as "core node" from the Qin dynasty to the Sui-Tang period in the calculations of *Wang, Ducruet and Wang* (2015). The supply of Chang´an on the contrast became more and more of a burden for the "imperial ecology"; its immediate hinterland was beset by drought and erosion frequently, diminishing crop yields. Already in the early 7$^{th}$ century, CE the court official Gao Jifu (d. 651) had criticised the Tang´s decision to establish residence there, since "*land is limited, and the people live densely together. Agriculture is not yielding. Beans and millet are cheap, but the stocks are not large*" (*Thilo* 2006, p. 183). So much more the maybe one to two million habitants of Chang´an and the many military garrisons nearby depended on the inflow of supplies via the canal network. Yet in contrast to Luoyang, Chang´an was not connected to the Grand Canal directly, but via the River Wei, which frequently changed his course and water level, respectively minor canals running parallel. The amount of grain, which could be transported to the capital, thus fluctuated between four million and 100,000 bushels. If supplies collapsed, the Tang emperors and the royal household had to relocate over 320 km to the east to Luoyang. This immensely expensive operation took place 14 times in the century between 640 and 740 CE, half of these occasions caused by supply shortfalls (*Thilo* 2006, pp. 193-194; *Thilo* 1997; *Elvin* 2004; *Lewis* 2009, p. 37; *von Glahn* 2016; *Xiong* 2017; *Preiser-Kapeller* 2018, pp. 237-238). Even when a new canal eased the situation from 743 CE onwards, the demands of the centre remained burdensome for the imperial ecology. Around that time, 36 % of all grain collected by the imperial administration and 48 % of all textiles paid as taxes had to be transported to Chang´an and its surrounding area. Like for imperial Rome, the urban metabolism of the Tang-capital almost entirely depended on the working of the tax- and distribution networks of the entire empire – and its less well-situated position as also reflected in the network model intensified the stress on the imperial ecology (*Xiong* 2006; *Thilo* 2006, 199-200; *Thilo* 1997; *von Glahn* 2016). Consequently, when the control of the Tang over the empire dwindled in the 9$^{th}$ century CE and made place for political fragmentation towards the end of that century, Chang´an shrank in scale and was officially abandoned in 904 CE. Characteristically, the court relocated to Luoyang, but also there the Tang rule ended in 907 CE. Only the Song succeeded in re-uniting most areas of China from 960 CE onwards; their new capital became Kaifeng (see **fig. 4**), which was even better located within the supply systems than Luoyang and surrounded by "*an intensive trade and communication zone*", as *William Guanglin Liu* (2015, p. 91) has stated. This is also reflected in our network analysis, where Kaifeng ranks fourth of all nodes in closeness and third in betweenness centrality (see **table 4**). However, when the Song lost the north of China to the Jin in 1126/1127 CE, they had to relocate their capital to Hangzhou further south. This place is equally well-connected in network analytical terms (ranking fourth of all nodes in degree in our model, see **table 4**) and near the open sea (see **fig. 8** and **fig. 10**), thus reflecting the increasing importance of maritime trade (which, as mentioned above, is unfortunately not reflected in our model) (*Brooke* 2014, pp. 347-348; *Thilo* 2006, pp. 24-28; *Twitchett* 1979, pp. 696, 720-728; Kuhn 2009, pp. 72-73, 224-227; *Tuan* 2008, pp. 132-135; *Mostern* 2011; *Schottenhammer* 2015; 2*Liu* 2015, pp. 77-95). The Mongols after their conquest of China (between 1235 and 1279 CE) established their capital at Khanbaliq/Dadu (modern-day Beijing), which was much more to the north than earlier capitals, but near the former capital Zhongdu of the Jin-Dynasty and to the regions of origins of the new Mongol Yuan-Dynasty (see **fig. 8** and **fig. 10**). When the Ming expelled the Yuan in 1368, the Ming kept Beijing as "Northern Capital" in addition to Nanjing as "Southern Capital" in the region where their rule has started. Both places became integrated into the Grand Canal network, which was extended towards the north; both sites are also well-connected in the network terms (see **fig. 8** and **fig. 10**), but Beijing ranks much higher in betweenness centrality (see **table 4**), also reflecting its strategic position at the routes towards the northern frontier, which became a permanent military challenge for the Ming (*Elvin* 1973; *Barfield* 1989; *Brook* 1998; *Brook* 2010; *Liu* 2015, pp. 106-120). In 1644, the Manchu coming from the Northeast captured the city and established the Ch'ing as last dynasty of imperial China (until 1911); they made Beijing China´s sole capital (*Huang* 1988, pp. 180-191; *Mote* 1999, pp. 813-911; *Peterson* 2002, pp. 563-640; *Elvin* 2004).

Following Carter, the application of network models confirms the idea that the position of capitals with the web of routes and corridors contributes to their emergence as centres and is, in turn, reinforced by the alignment of infrastructures on their demands. Increased connectivity within the imperial ecology had, however, also unintended consequences, such as the facilitation of the diffusion of epidemics. Under the early Tang, a major contagion between 636 and 643 spread from Chang´an to the east already primarily along the recently established axes of the Grand Canal System, reaching Luoyang and Kaifeng (see **fig. 11**). Equally, epidemics under the Ming in 1588 and in 1642 CE (see **fig. 12** and **fig. 13**) followed the main corridors of connectivity (by closeness centrality) identified in our network analysis (*Elvin* 2004; *Marks* 2017, pp. 146-148). For the three major epidemics registered in Roman imperial times (the Antonine plague in 165-180 CE, Cyprian´s Plague in 249-262 CE and Justinian´s Plague, whose waves afflicted the Mediterranean between 540 and 750 CE), unfortunately the data on their spatial diffusion is much more sparse. However, for the first outbreak of Justinian´s Plague in the 540s, probable corridors of spread show equally a high overlap with the zones of high closeness-centrality marked in our analysis of the ORBIS-network (see **fig. 14**). Imperial ecologies thus also became disease ecologies, the later using the networks of the former (*Bianconi* 2018, pp. 58-66; *Cliff and Smallman-Raynor* 2009; *McCormick* 2003; *Stathakopoulos* 2004; *Little* 2006; *Harper* 2017).

The city of Rome, however, had already experienced a significant reduction of size by the mid-6th century, which had not been caused by the plague; as Baccini and Brunner underline, "*the drastic shrinking was not due to an ecological collapse but to an institutional breakdown. The metabolism of such large systems is not robust because it cannot maintain itself without a huge colonized hinterland. It has to reduce its population to a size that is in balance with its economically and ecologically defined hinterland.*" (*Baccini and Brunner* 2012, p. 58) The urban shrinking of Rome was one consequence of the breakdown of central rule and the fragmentation of the imperial networks (and ecology) in the western provinces in the 5th century CE. This process is traditionally marked with the sacks of Rome in 410 and 455 CE, the loss of the vital breadbasket of North Africa to the Vandals (between 429 and 439 CE) and finally the dismissal of Emperor Romulus Augustulus in 476 CE (who characteristically had his residence in Ravenna) (*Börm* 2013; *Preiser-Kapeller* 2016; *Preiser-Kapeller* 2018, pp. 224-225). This leads to the question of the robustness and the possible dynamics of fragmentation of imperial networks.

**Robustness and fragmentation of imperial networks**

As discussed above, complex networks are not uniformly connected; we have observed big differences in centrality measures between nodes. Equally, networks are often structured in clusters, meaning groups of nodes, which are more densely and closely connected among each other than with the rest of the network; they may be identified as "sub-communities" within the larger system. For their identification, one can use algorithms for "group detection", such as the one developed by the physicist *M. Newman* (2010, pp. 372-382), which aims at an "optimal" partition of the network into clusters. Complex network are as well characterised by "nested clustering", such that within clusters further sub-clusters can be detected, within which further cluster can be identified, across several levels of hierarchy (Barabási 2016, pp. 331-338).

For the ORBIS-model, with the help of the Newman-algorithm we identified 25 over-regional clusters of higher internal connectivity (see **table 5** and **fig. 15**). The majority of these clusters owe their connectivity to either maritime connections (nr. 5, 6, 7, 8, 9, 12, 13, 15, 16, 17, 18,

19, 20, 21, 22, 24, 25) or riverine routes (nr. 1, 3, 4, 14, 23) (see also *McCormick* 2001, pp. 77-114). In order to test the concept of "nested clustering", we applied the Newman-algorithm also on each of the 25 (over)regional clusters of the ORBIS-network, resulting in the identification of between three and eight regional sub-clusters within each of the larger clusters (see **fig. 16**). We may therefore perceive this complex network model of localities and routes in the Roman Empire across several spatial scales as a system of nested clusters, down to the level of individual settlements and their hinterlands (on the Mediterranean as "agglomeration" of micro-regions and the role of (imperial) connectivity see also *Horden and Purcell* 2000; *Manning* 2018). In such a network, speed and cohesion of empire-wide connectivity depends on the trans-regional links between these clusters, which structure the entire system.

A similar picture emerges if we apply the same clustering-algorithm of Newman on the network model for China (see **table 6** and **fig. 17**). Especially the major clusters (such as nrs 1, 10, 14, 19, 23) emerge based on riverine connectivity, while the largest cluster nr 3 (see **fig. 17**) connects the places along the Grand Canal network (also including the four ancient capitals of Luoyang, Kaifeng, Nanjing and Hangzhou). These (like in the Roman case) especially "hydrous", over-regional linkages allow for a cohesion of the imperial network at large and its integration of macro- and micro-regions into one overarching system (for the actual regional structure of imperial China cf. also *Mostern* 2011).

The results of the Newman clustering algorithm, however, are only one of various possible solutions to the problem of community detection. Different clustering algorithms will produce different attributions of nodes into clusters, such as the Louvain algorithm, which we also applied on both network models (for an example, see **fig. 18**). Even the same Newman algorithm will propose a different partition into clusters of the same network if some parameters (such as relative link weights of land, riverine and maritime connections) are modified. Yet, all results for the two network models show the same pattern of nested clusters from the over-regional down to the local level, following the same "logics" of increased connectivity (trans-maritime linkages in the Roman case, for instance) (*Newman* 2010, pp. 354-392; *Estrada* 2012, pp. 187-213; *Barabási* 2016, pp. 320-362).

However, could these boundaries between clusters also work as potential rupture lines in case of a weakening of the network´s cohesion? The robustness respectively vulnerability of complex networks has attracted a considerable amount of attention, not the least due to the relevance of these questions for modern-day infrastructural webs. As *Ginestra Bianconi* (2018, pp. 49-57) has stated, "*it is assumed that a fundamental proxy for the proper function of a given* *network is the existence of a giant component, (…) which allows the propagation of ideas, information and signals*" as well as resources and people across (most of) the network. The extent to which such as giant component within a given network exists is indicated by the so-called "percolation threshold", which allows for the existence of large clusters and long-range connectivity. Below this threshold, a network disintegrates in various components of smaller size, and system-wide connectivity is severely damaged (see also *Wang and Ducruet* 2013). One test of a network´s robustness is the successive removal of nodes, "*monitoring the fraction of nodes that remains in the giant component of the network after the inflicted damage*", thus "simulating" a cascading failure or destruction of nodes. The removal of nodes can be executed randomly or in the form of a "targeted attack", when nodes are damaged "*according to a non-random strategy*" such as selecting nodes which rank high in certain centrality measures. Characteristically, large-scale complex networks would be very robust vis-à-vis random attacks, since due to the inequality in the distribution of centrality values (see **fig. 1** and **fig. 2**), there is a high probability that only peripheral nodes are affected while the central hubs remain intact. A targeted attach on the latter, however, could lead to a rapid fragmentation of the network (*Bianconi* 2018, pp. 49-57). We used the targeted attack approach and successively removed the nodes ranking on top in betweenness centrality until a considerable share (at least 20 %) of the network was no longer connected to the original giant component.

Both network models show a significant robustness even towards such a "non-random strategy": in the ORBIS-network, we removed the top 50 nodes in betweenness (that is 7.3 % of all nodes of the unmodified network) before we observed a major disruption (see **fig. 19**). Interestingly, after this rupture, the North-western regions of the network emerge as separate component (nr 1, in green colour), while the entire Mediterranean area is integrated in one (still relatively giant) component (nr 2, in red colour) (see **fig. 19**). This again highlights the significance of maritime connectivity for the network´s cohesion, but (very tentatively) could also be connected to post-476 CE scenarios of an attempted renovation of Mediterranean imperial unity by Emperor Justinian in the 6$^{th}$ century CE or the (now of course out-dated) "Pirenne-thesis" about the emancipation of the Frankish Empire from the Mediterranean core (*Preiser-Kapeller* 2018).

The China-network proves to be even more robust to targeted node failure. Only after a removal of the 150 top nodes in betweenness (14.5 % of all nodes of the unmodified network), smaller separate components emerge in the Northeast (nr 2), the Northwest (nr 7) and especially in the South (nr 4), while the core at large remains intact (component nr 1, in orange colour), including all traditional imperial capitals (see **fig. 20**). Again, we attribute this to the cohesive effect of the riverine connections, augmented by large-scale imperial infrastructural projects such as the Grand Canal.

Yet, what happens, if these relatively cost-intensive, maybe even "fragile links" across larger distances, as *Ward-Perkins* (2006, p. 382) has called them for the Roman Empire, "disappear"? In order to answer this question, we applied another robustness text and eliminated step by step all links from the network models above a specific "cost" threshold; this could be interpreted as a "simulation" of the dwindling ability of an imperial centre to maintain (or defend) expensive and vulnerable long-distance connections and infrastructures.

From the ORBIS-network, we successively removed all links which would "cost" more than five, more than three, more than two and finally more than one day´s journey(s) (according to the calculations of the ORBIS-team) (see **table 1**). The result is an increasing fragmentation of the network in components of different size, partially along the "rupture lines" between the clusters and sub-clusters, which we identified for the unmodified network model (see above). But even if we eliminate the connections across longer distances, some larger, over-regional clusters, again especially of maritime connectivity, demonstrate remarkable robustness (see also *McCormick* 2001, pp. 565-569, on the resilience of certain sea routes in the $7^{th}$ to $9^{th}$ cent. CE). In the model, in which all connections, which "cost" more than one day´s journey, are deleted (see **table 1** and **fig. 21**), the largest still fully connected component (nr 6, in yellow colour) is located in the Eastern Mediterranean between the Tyrrhenian Sea and the Levant, with its centre in the Aegean. This would correspond to the central regions and communication routes, which remained under control of the (Eastern) Roman Empire after the loss its eastern provinces to the Arabs in the $7^{th}$ century CE, at the end of an actual process of increasing fragmentation of the (post)Roman world (*Brubaker and Haldon* 2011; *Vaccaro* 2013; *Haldon* 2016). We applied the Newman-algorithm in turn on this remaining largest component (nr 6) and identified again various regional clusters nested within the larger connected system, especially in the Aegean and along the coasts of Asia Minor; interestingly, also the (former) imperial residences of Rome and Ravenna "have" resilient medium-sized components intact in this scenario (see **fig. 22**). Yet besides the resilience of maritime connectivity in regions of Italy and the Eastern Mediterranean (and equally an uninterrupted cohesion of the "Egyptian" cluster, see also *Wickham* 2005, pp. 759-769), we observe a general "disentanglement" of large parts of the Roman traffic system, especially in the West of Europe, equally in the interior of the Balkans or also between the North and Souths coasts of the Mediterranean (see **table 1** and **fig. 21**). The model is of course at best an appropriation towards certain structural parameters of the web of transport links within the Imperium Romanum. Nevertheless, we observe some remarkable parallels to actual historical processes of the 5th to 7th century CE (*Wickham* 2004 for instance wrote about a partial "*micro-regionalisation*" of the "*Mediterranean world-system*" during this period), which hint at the impact of processes of integration respectively disentanglement especially due to the establishment and growth respectively the contraction of long distance connections (*McCormick* 2001, pp. 270-277, 385-387).

We executed the same test on the China-network model, removing all links, which would "cost" more than five, more than three, more than two and finally more than one day´s journey(s) (see **table 2**). In this case, a major impact on the connectedness can be observed after the removal of all links "worth" more than two days of travel (see **fig. 23**), with smaller components emerging in the West (nr 5, with Chang´an), Northwest (nr 27), Northeast (nr 3), South (nr 2) and Southwest (nr 16) (on actual regional faults in Chinese history cf. *Schmidt-Glintzer* 1997). Still, one major component covers the entire east of the network, including the central region along the Grand Canal(s) with the capitals of Beijing, Luoyang, Kaifeng, Nanjing and Hangzhou and ranging all the way to the South until Fujian province (with a total of 478 nodes or 46 % of the unmodified network) (see **fig. 23**). The next step, however, i.e. the elimination of all links "costing" more than one day of journey, results in a total fragmentation of the model, with a multitude of small-scale networks none containing more than 20 nodes (or 0.02 % of the original network) (see **table 2** and **fig. 24**). In contrast to the ORBIS-model, no resilient larger component emerges after this robustness test. This may indicate that from a structural point of view imperial cohesion over larger territories in China came at a greater cost than in the Mediterranean, maritime-based network. Against such a scenario, the relative endurance of unified imperial regimes in China in contrast to the more fragmented (and after the 5th century CE never again entirely politically integrated) Mediterranean is even more remarkable (cf. also *Schmidt-Glintzer* 1997; *Scheidel* 2009).

**Conclusion**

The actually very different historical trajectories of the Euro-Mediterranean region and of China rebut any deterministic interpretation of a structural-quantitative approach on empires of the past (see also *Scheidel* 2009). Although also recent studies suggest a long term impact of the imperial infrastructures of Rome or ancient China even on modern-day economic performance (*Fang, Feinman and Nicholas* 2015; *Dalgaard, Kaarsen, Olsson and Selaya* 2018), understanding them as complex networks leads to an expectation a high diversity of responses to internal dynamics and external challenges, especially across spatial scales. Remarkable resilience at the regional level can take place at the same time when the system at large disintegrates; the multitude of developments in the "post-Roman" world as highlighted in recent research would conform with such a "complex behaviour" (*Cameron, Ward-Perkins and Whitby* 2000; *Sarris* 2011; *Demandt* 2015; *Preiser-Kapeller* 2016). Equally, in the Chinese case, imperial unity was no "frozen evolutionary path", as periods of political multiplicity in the 4$^{th}$-6$^{th}$ century, in the 10$^{th}$ century or also in the first half of the 20$^{th}$ century CE after the fall of the Ch'ing dynasty indicate (*Huang* 1988; *Elvin* 1973). On the other hand, China could serve as example also of the relative robustness of large-scale imperial networks at large under changing regimes and after episodes of fragmentation. The establishment of the Grand Canal network also initiated a certain "path dependence" with regard to the selection of "nodes" as centres of the imperial system. When it comes to network analytical measures, Chinese rulers were rather "successful" in their decision-making if we follow F. W. Carter again. Based on our findings, we may also confirm and conclude with his verdict, that "*there therefore seems little excuse why the historical geographer should not attempt to use some of these techniques in his analysis of certain aspects of the past.*" (*Carter* 1969, p. 46)

**Tables**

| ORBIS network model | Unmod. network | W/o links of more than 5 d of travel | more than 3 d of travel | more than 2 d of travel | more than 1 d of travel |
|---|---|---|---|---|---|
| **Number of nodes** | 678 | 678 | 678 | 678 | 678 |
| **Number of edges** | 1104 | 1005 | 870 | 673 | 424 |
| **Connectedness** | 1 | 0,93 | 0,589 | 0,325 | 0,053 |
| **Number of isolates** | 0 | 13 | 50 | 117 | 246 |
| **Density** | 0,005 | 0,004 | 0,004 | 0,003 | 0,002 |
| **Diffusion (reach of information across the network)** | 0,986 | 0,894 | 0,461 | 0,256 | 0,029 |
| **Clustering Coefficient (largest component)** | 0,116 | 0,12 | 0,137 | 0,126 | 0,131 |
| **Fragmentation** | 0 | 0,07 | 0,411 | 0,675 | 0,947 |
| **Betweenness Centralisation** | 0,504 | 0,336 | 0,216 | 0,145 | 0,013 |
| **Circuitry (Alpha-index)** | 0,32 | 0,24 | 0,14 | 0,00 | -0,19 |
| **Size of largest component** | 678 | 640 | 519 | 385 | 148 |
| **Av. range of travel between nodes (biggest component, unmod. = 1)** | 1,00 | 0,90 | 0,76 | 0,61 | 0,41 |

**Table 1:** Network measures for the unmodified ORBIS-network model and the four modified models with removed links above cost-thresholds

| China network model | Unmod. network | W/o links of more than 5 d of travel | more than 3 d of travel | more than 2 d of travel | more than 1 d of travel |
|---|---|---|---|---|---|
| **Number of nodes** | 1034 | 1034 | 1034 | 1034 | 1034 |
| **Number of edges** | 3034 | 2788 | 2056 | 1029 | 315 |
| **Connectedness** | 1 | 0,985 | 0,935 | 0,254 | 0,003 |
| **Number of isolates** | 0 | 2 | 5 | 37 | 353 |
| **Density** | 0,011 | 0,01 | 0,008 | 0,004 | 0,001 |
| **Diffusion (reach of information across the network)** | 1 | 0,975 | 0,922 | 0,248 | 0,003 |

| Clustering Coefficient (largest component) | 0,65 | 0,643 | 0,62 | 0,478 | 0,133 |
|---|---|---|---|---|---|
| Fragmentation | 0 | 0,015 | 0,065 | 0,746 | 0,997 |
| Betweenness Centralisation | 0,173 | 0,395 | 0,302 | 0,123 | 0 |
| Circuitry (Alpha-index) | 0,97 | 0,85 | 0,50 | 0,00 | -0,35 |
| Size of largest component (number of nodes) | 1034 | 1026 | 1000 | 478 | 20 |
| Av. range of travel between nodes (biggest component, unmod. = 1) | 1 | 0,9 | 0,75 | 0,5 | 0,272 |

**Table 2:** Network measures for the unmodified China-network model and the four modified models with removed links above cost-thresholds

| Cities (in alphabetic order) | Degree [scaled, network mean = 1] (rank) | Betweenness [scaled, network mean = 1] (rank) | Closeness [scaled, network mean = 1] (rank) |
|---|---|---|---|
| Antioch | 0,35 | 0,29 | 0,99 |
| Aquileia | 2,92 | 5,38 | 1,29 |
| Constantinople | 6,70 (**9**) | 6,61 (**10**) | 1,34 (**10**) |
| Milan | 0,24 | 4,82 | 1,31 |
| Nicomedia | 2,04 | 0,34 | 1,08 |
| Ravenna | 2,32 | 0,03 | 1,11 |
| Rome | 0,85 | 4,33 | 1,11 |
| Serdica | 0,02 | 9,23 (**4**) | 1,35 (**4**) |
| Sirmium | 0,11 | 4,11 | 1,30 |
| Thessalonike | 0,92 | 3,22 | 1,24 |
| Trier | 0,15 | 0,21 | 1,14 |

**Table 3:** Scaled node centrality measures for selected Roman imperial residential cities in the ORBIS-network model (positions of a city within centrality rankings for all nodes of the network are only indicated if among the top 20)

| Cities (in alphabetic order) | Degree [scaled, network mean = 1] (rank) | Betweenness [scaled, network mean = 1] (rank) | Closeness [scaled, network mean = 1] (rank) |
|---|---|---|---|
| Beijing | 2,28 | 11,91 (**11**) | 1,0166 |
| Chang´an | 0,74 | 11,16 (**13**) | 1,0109 |
| Hangzhou | 4,77 (**4**) | 3,85 | 1,0187 |
| Kaifeng | 3,60 (**12**) | 18,38 (**3**) | 1,0208 (**4**) |
| Luoyang | 3,50 (**13**) | 11,52 (**12**) | 1,0206 (**8**) |
| Nanjing | 2,86 | 4,38 | 1,0195 (**18**) |

**Table 4:** Scaled node centrality measures for selected Chinese imperial residential cities in the China-network model (positions of a city within centrality rankings for all nodes of the network are only indicated if among the top 20)

| Newman-Cluster nr. | Regions of the Roman Empire |
|---|---|
| 1 | Upper Danube, Eastern Alps |
| 2 | North Syria, Northwest Mesopotamia, South Asia Minor |
| 3 | Rhine area |
| 4 | Middle Danube, North Balkans |
| 5 | North and central Adriatic |
| 6 | Central North Africa, East coast of Iberian Peninsula, Baleares |
| 7 | South Iberian Peninsula, West North Africa |
| 8 | Region around the Sea of Marmara, North Aegean |
| 9 | Central and Northwest Aegean |
| 10 | Central South Italy |
| 11 | Palestine |
| 12 | Britannia and Channel coast |
| 13 | Rome, Latium and Campania |
| 14 | Egypt |
| 15 | Cyrenaica, Crete and South Peloponnese |
| 16 | Southwest Asia Minor |
| 17 | Cyprus and north coasts of Levant |
| 18 | East North Africa, Sicily and Southwest of South Italy |
| 19 | Black Sea and North of Asia Minor |
| 20 | Etruria, Liguria, Corsica and Southeast of Gaul |
| 21 | Gaul and North of the Iberian Peninsula |
| 22 | South Adriatic and North Epirus |

| 23 | Western plain of the river Po |
| 24 | Northwest central Greece, North Peloponnese and Ionian Sea |
| 25 | Central Aegean (micro-cluster) |

**Table 5:** Regional attributions of clusters of nodes in the ORBIS-network model identified with the help of the Newman-algorithm

| Newman-Cluster nr. | Regions and provinces of contemporary China |
|---|---|
| 1 | Shaanxi province with the ancient capital of Chang'an |
| 2 | Hebei province and the Beijing area |
| 3 | The regions along the Grand Canals from Hangzhou up to Hebei |
| 4 | The east of Guangdong province |
| 5 | Parts of Anhui and Jiangxi provinces |
| 6 | Parts of Liaoning province |
| 7 | The west of Guangdong and parts of Hunan province |
| 8 | Shandong province |
| 9 | Fujian province |
| 10 | The central core between Chang'an in the north to Hunan in the south |
| 11 | Coastal parts of Zhejiang province |
| 12 | Parts of Sichuan province |
| 13 | Guangxi province |
| 14 | Parts of Sichuan and Chongqing provinces |
| 15 | Guizhou province |
| 16 | Eastern parts of Gansu province |
| 17 | Main parts of Gansu province |
| 18 | Southernmost parts of Guangdong province |
| 19 | Yunnan province (major part) and adjacent regions |
| 20 | Shanxi province |
| 21 | Micro-cluster in western Guangxi province |
| 22 | Westernmost parts of Yunnan province |
| 23 | Jiangxi province and adjacent areas |
| 24 | Ningxia province |
| 25 | Southernmost parts of Sichuan province |
| 26 | Easternmost parts of Yunnan province |

**Table 6:** Regional attributions of clusters of nodes in the China-network model identified with the help of the Newman-algorithm

*Marks, Robert B. (2017)*: China. An Environmental History. – Lanham, Boulder, New York and London.

*McCormick, Michael (2001)*: Origins of the European Economy. Communications and Commerce AD 300-900. – Cambridge.

*McCormick, Michael (2003)*: Rats, Communications, and Plague. Towards an Ecological History. – In: Journal of Interdisciplinary History 34/1, 2003, pp. 1-25.

*Morley, Neville (1996)*: Metropolis and Hinterland. The City of Rome and the Italian Economy 200 BC-AD 200. – Cambridge.

*Mostern, Ruth (2011)*: "Dividing the Realm in Order to Govern". The Spatial Organization of the Song State (960-1276 CE). – Cambridge, Mass. and London.

*Mote, Frederick W. (1999)*: Imperial China 900-1800. – Cambridge, Mass. and London.

*Mutschler, Fritz-Heiner and Mittag, Achim (eds.) (2008)*: Conceiving the Empire. China and Rome Compared. – Oxford.

*Newman, M. (2010)*: Networks. An Introduction. – Oxford.

*Orengo, Hector A. and Livarda, Alexandra (2016)*: The seeds of commerce: a network analysis-based approach to the Romano-British transport system. – In: Journal of Archaeological Science 66, 2016, pp. 21-35.

*Peterson, Willard J. (ed.) (2002)*: The Cambridge History of China, Vol. 9. Part One: The Ch'ing Empire to 1800. – Cambridge.

*Pfeilschifter, Rene (2014)*: Die Spätantike: Der eine Gott und die vielen Herrscher. – Munich.

*Pitts, F. R. (1978)*: The Medieval River Trade Network of Russia Revisited. – In: Social Networks 1, 1978, pp. 285-292.

*Preiser-Kapeller, Johannes (2015)*: Harbours and Maritime Networks as Complex Adaptive Systems – a thematic Introduction. –In: Preiser-Kapeller, Johannes and Daim, Falko (eds.), Harbours and Maritime Networks as Complex Adaptive Systems. Mainz, pp. 1-23.

*Preiser-Kapeller, Johannes (2016)*: Byzantinische Geschichte, 395-602. – In: Daim, Falko (ed.), Byzanz. Historisch-kulturwissenschaftliches Handbuch, Stuttgart 2016, pp. 1-61.

*Preiser-Kapeller, Johannes (2018)*: Jenseits von Rom und Karl dem Großen. Aspekte der globalen Verflechtung in der langen Spätantike, 300-800 n. Chr. – Vienna.

*Preiser-Kapeller, Johannes (2019)*: Networks as Proxies: a relational approach towards economic complexity in the Roman period. – In: Verboven, Koen and Poblome, Jeroen (eds.), Structure and Performance in the Roman Economy: Complexity Economics. Finding a New Approach to Ancient Proxy Data. forthcoming 2019.

*Prell, Christina (2012)*: Social Network Analysis. History, Theory and Methodology. – Los Angeles and London.

*Rodrigue, Jean-Paul; Comtoi, Claude and Slack, Brian (2013)*: The Geography of Transport Systems. – 3rd ed., London and New York.

*Ruffing, Kai (2012)*: Wirtschaft in der griechisch-römischen Antike. – Darmstadt.

*Sarris, Peter (2011)*: Empires of Faith. The Fall of Rome to the Rise of Islam, 500-700. – Cambridge.

**Figures**

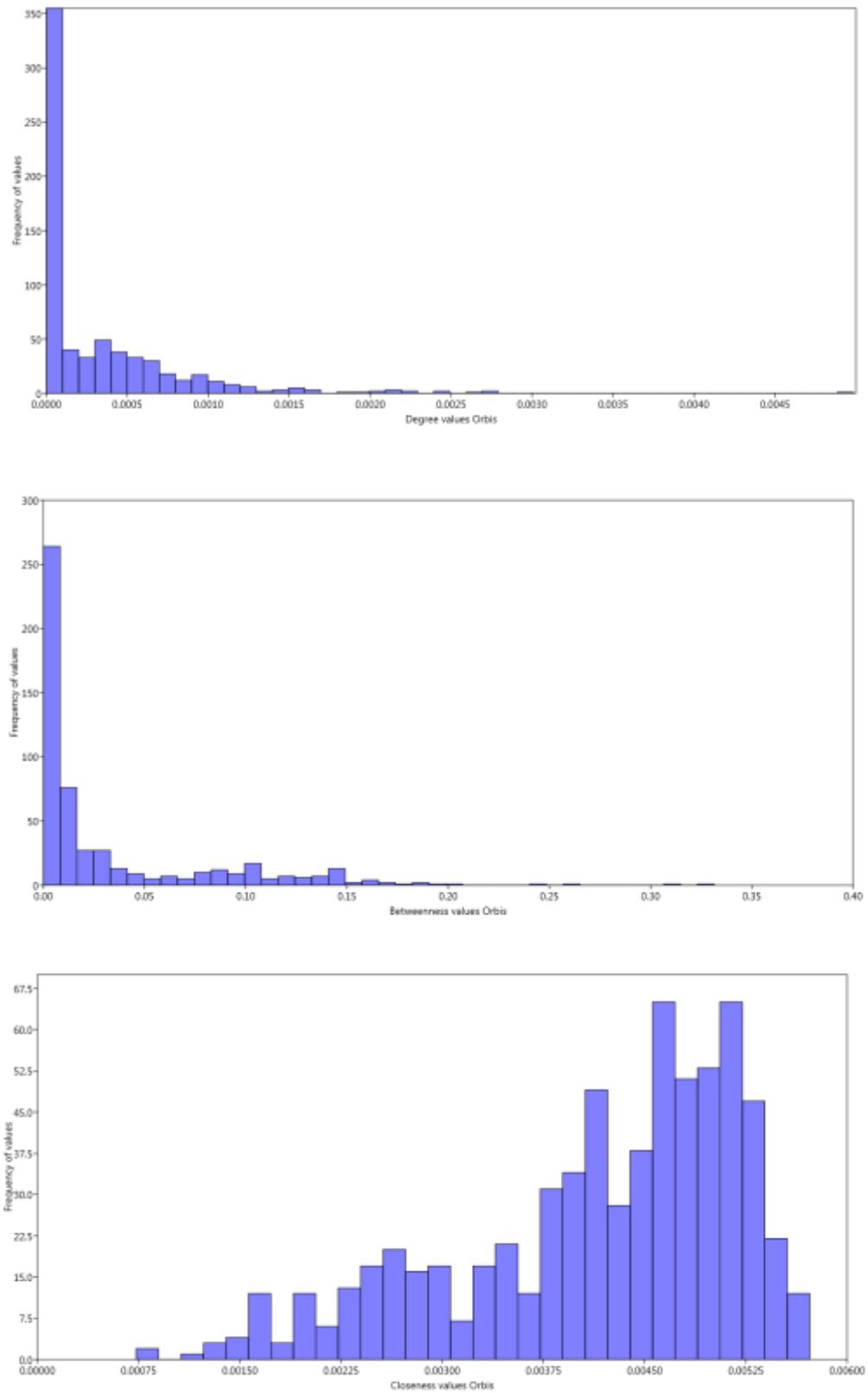

**Fig. 1:** Histograms of the distribution of degree values (top), betweenness values (centre) and closeness values (bottom) among all nodes in the ORBIS-network model (graph: J. Preiser-Kapeller, 2018)

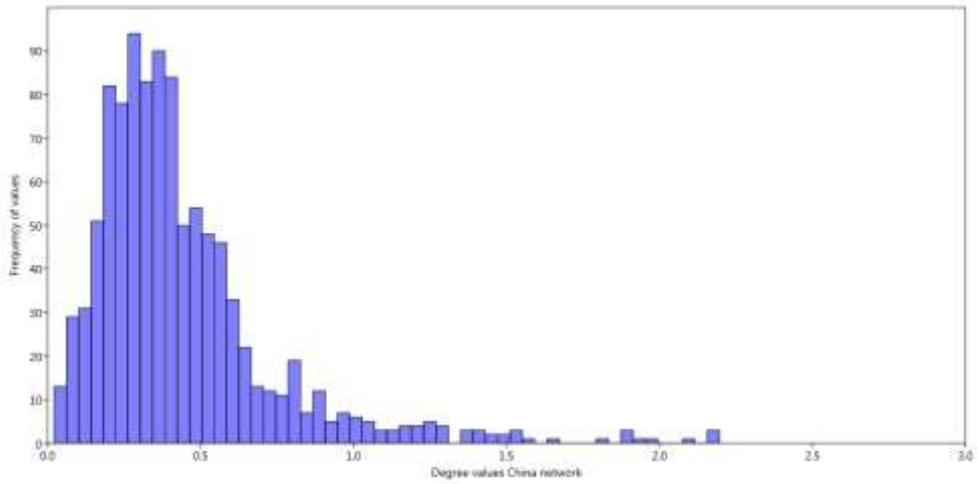

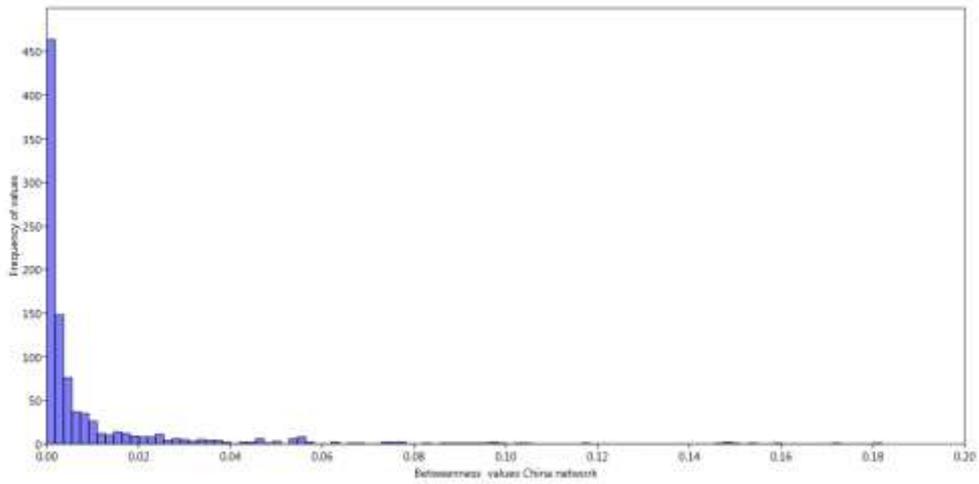

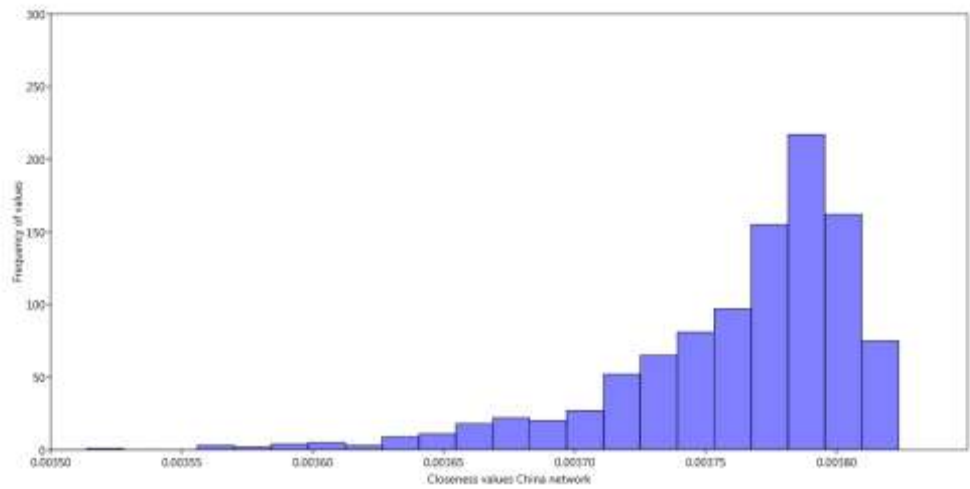

**Fig. 2:** Histograms of the distribution of degree values (top), betweenness values (centre) and closeness values (bottom) among all nodes in the China-network model (graph: J. Preiser-Kapeller, 2018)

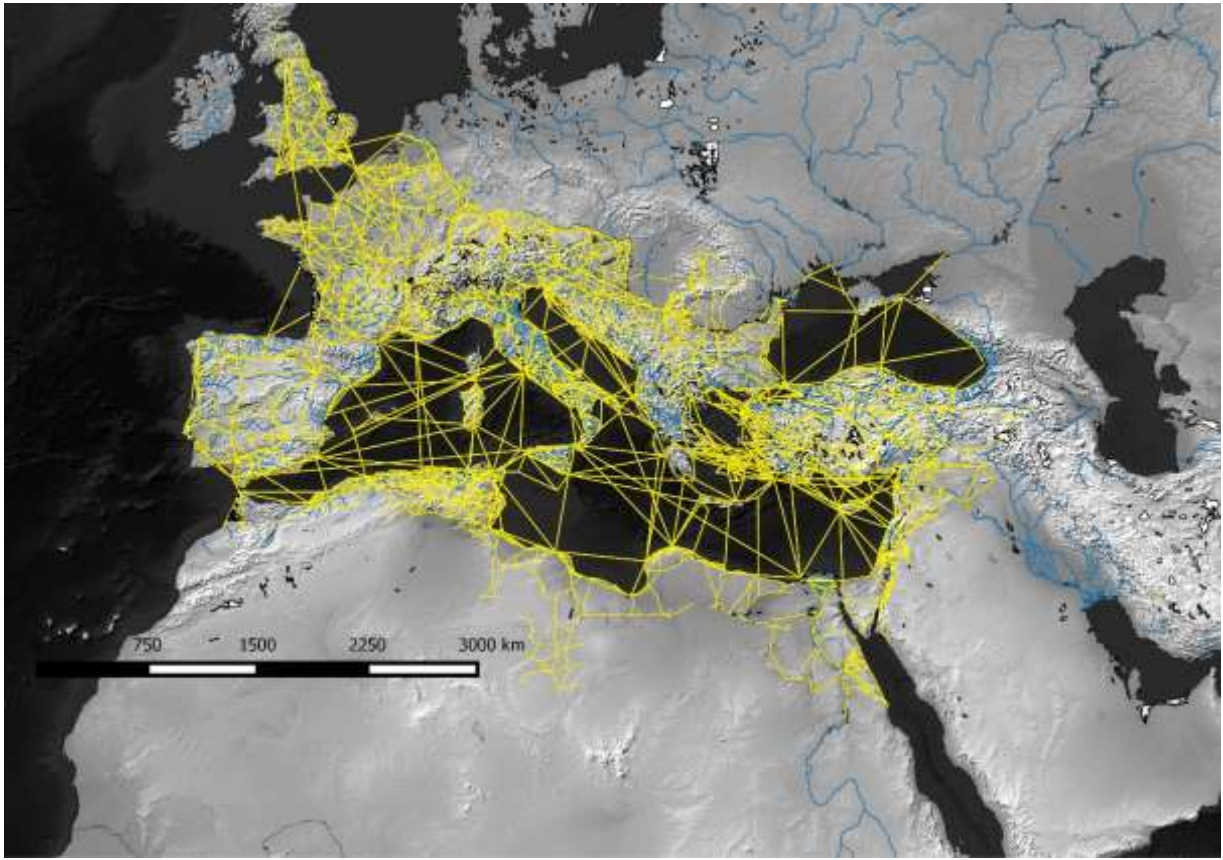

**Fig. 3:** Terrestrial, riverine and maritime routes in the ORBIS-network model for the Roman Empire (data: http://orbis.stanford.edu/; map: J. Preiser-Kapeller, 2018)

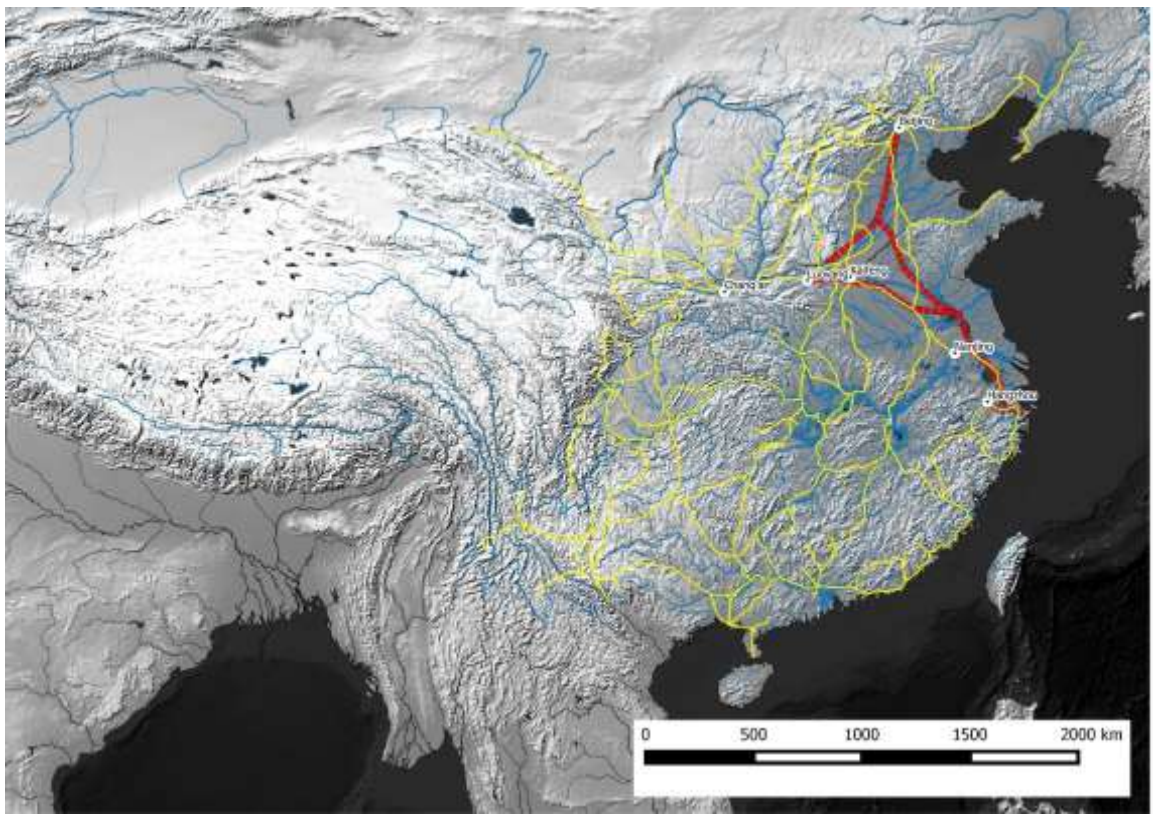

**Fig. 4:** Terrestrial, riverine and canal routes in the network model for Imperial China (red = Grand Canal system; data: http://sites.fas.harvard.edu/~chgis/; map: J. Preiser-Kapeller, 2018)

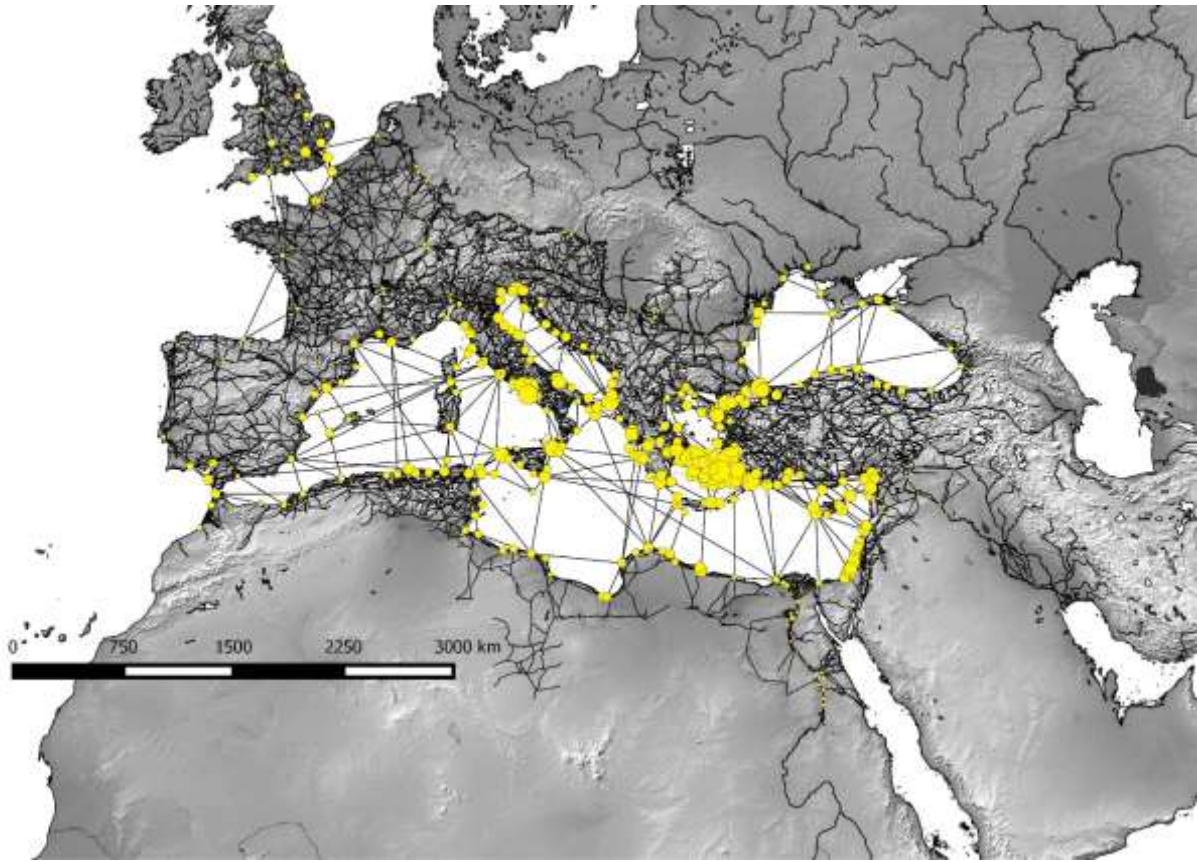

**Fig. 5:** Spatial distribution of degree values of nodes in the ORBIS-network model for the Roman Empire (data: http://orbis.stanford.edu/; calculations and map: J. Preiser-Kapeller, 2018)

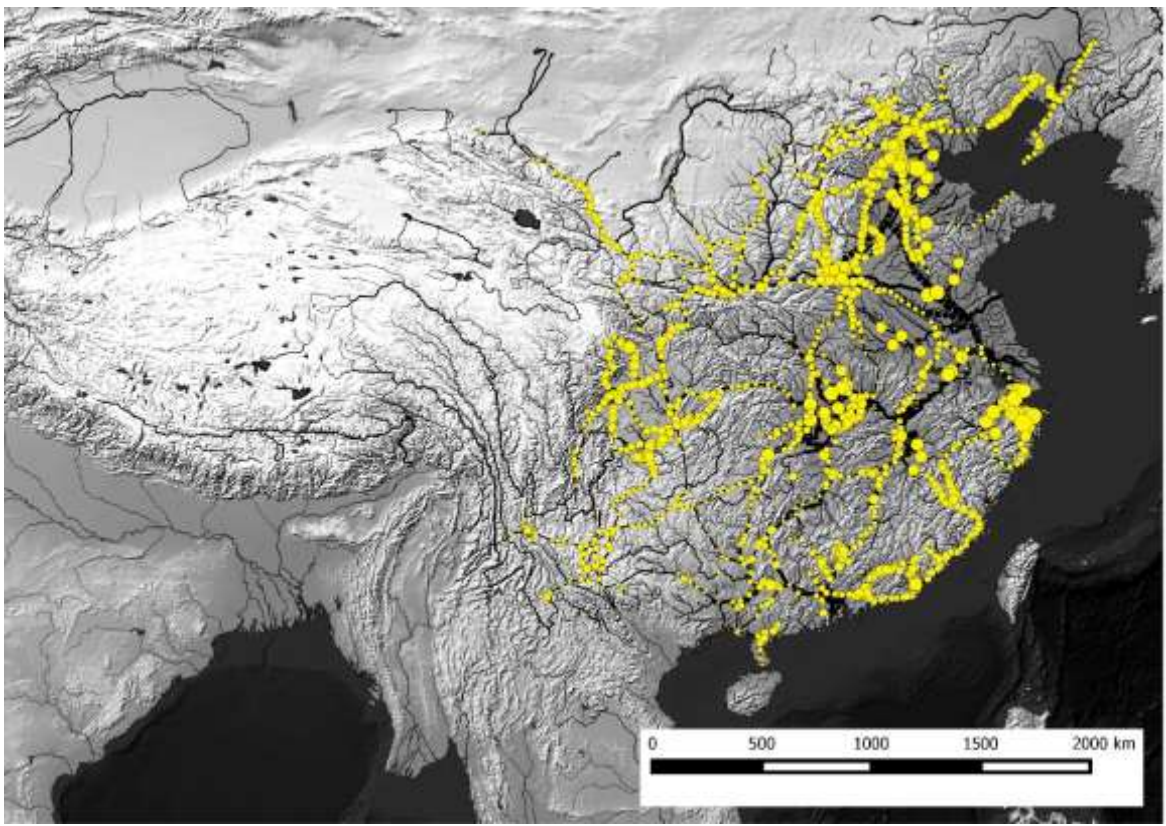

**Fig. 6:** Spatial distribution of degree values of nodes in the network model for Imperial China (data: http://sites.fas.harvard.edu/~chgis/; calculations and map: J. Preiser-Kapeller, 2018)

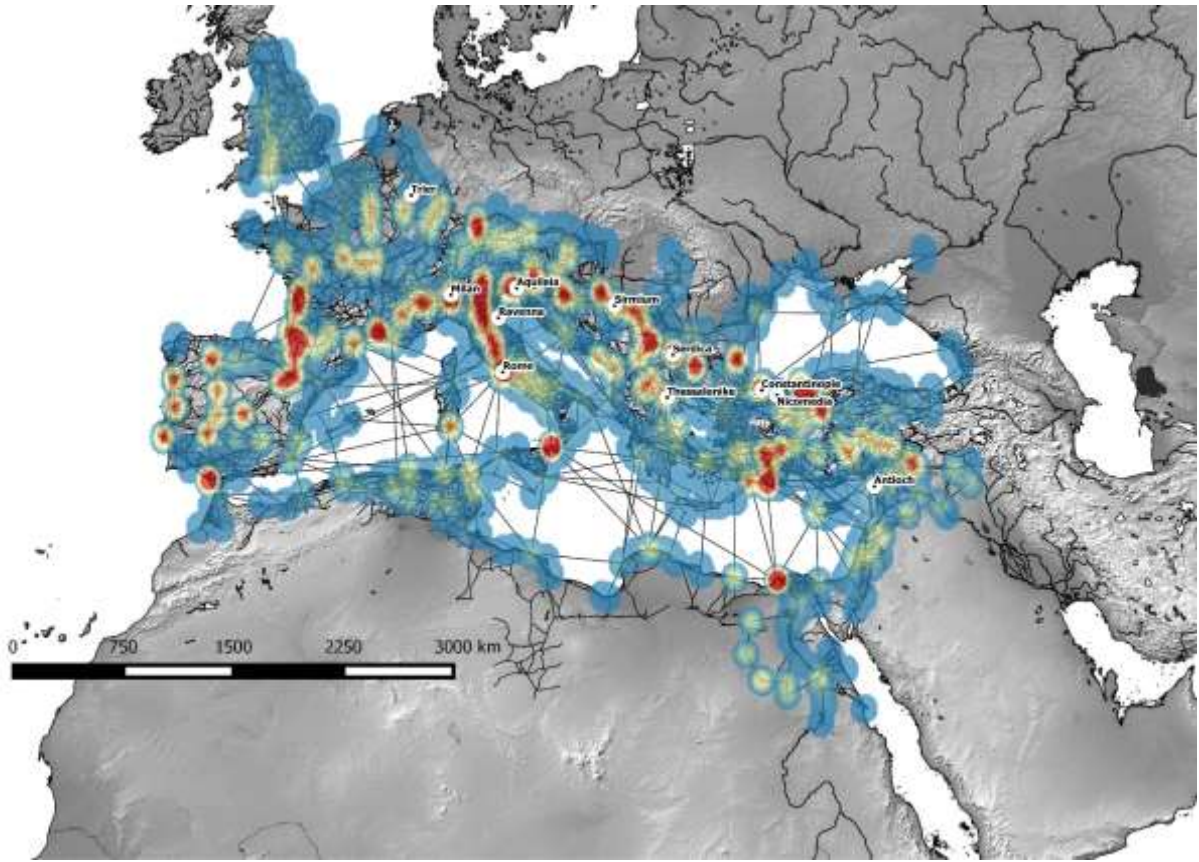

**Fig. 7:** Heatmap of the spatial distribution of betweenness values of nodes in the ORBIS-network model for the Roman Empire (data: http://orbis.stanford.edu/; calculations and map: J. Preiser-Kapeller, 2018)

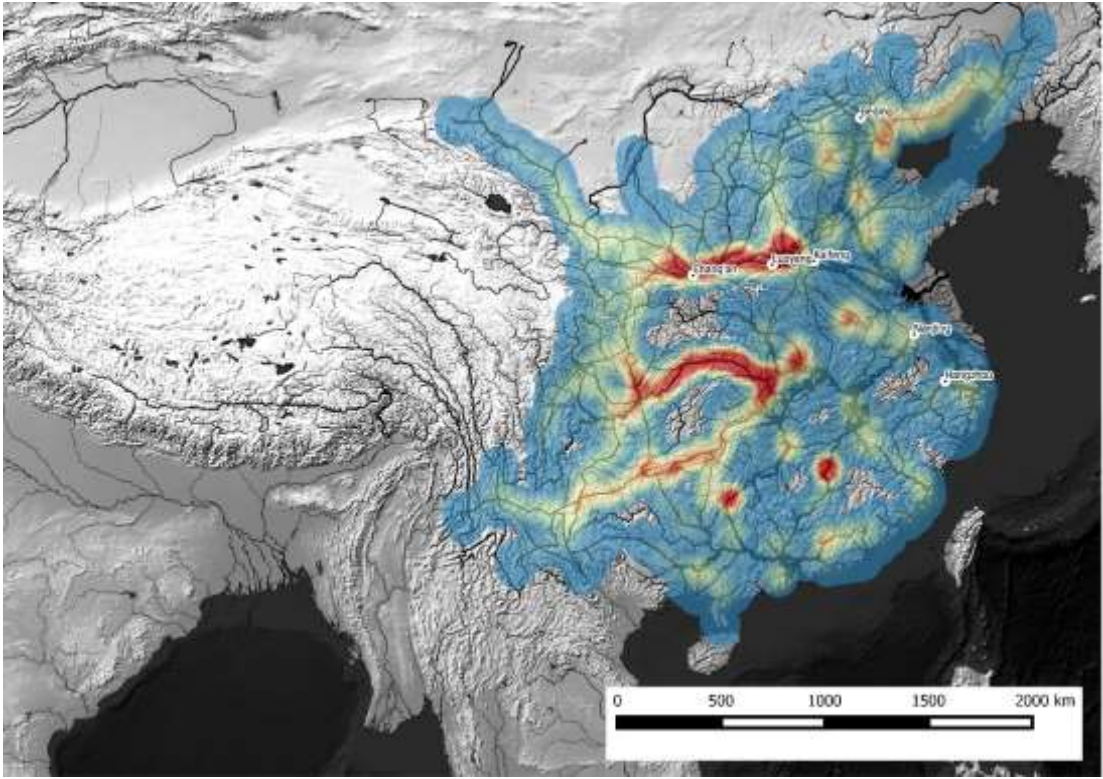

**Fig. 8:** Heatmap of the spatial distribution of betweenness values of nodes in the network model for Imperial China (data: http://sites.fas.harvard.edu/~chgis/; calculations and map: J. Preiser-Kapeller, 2018)

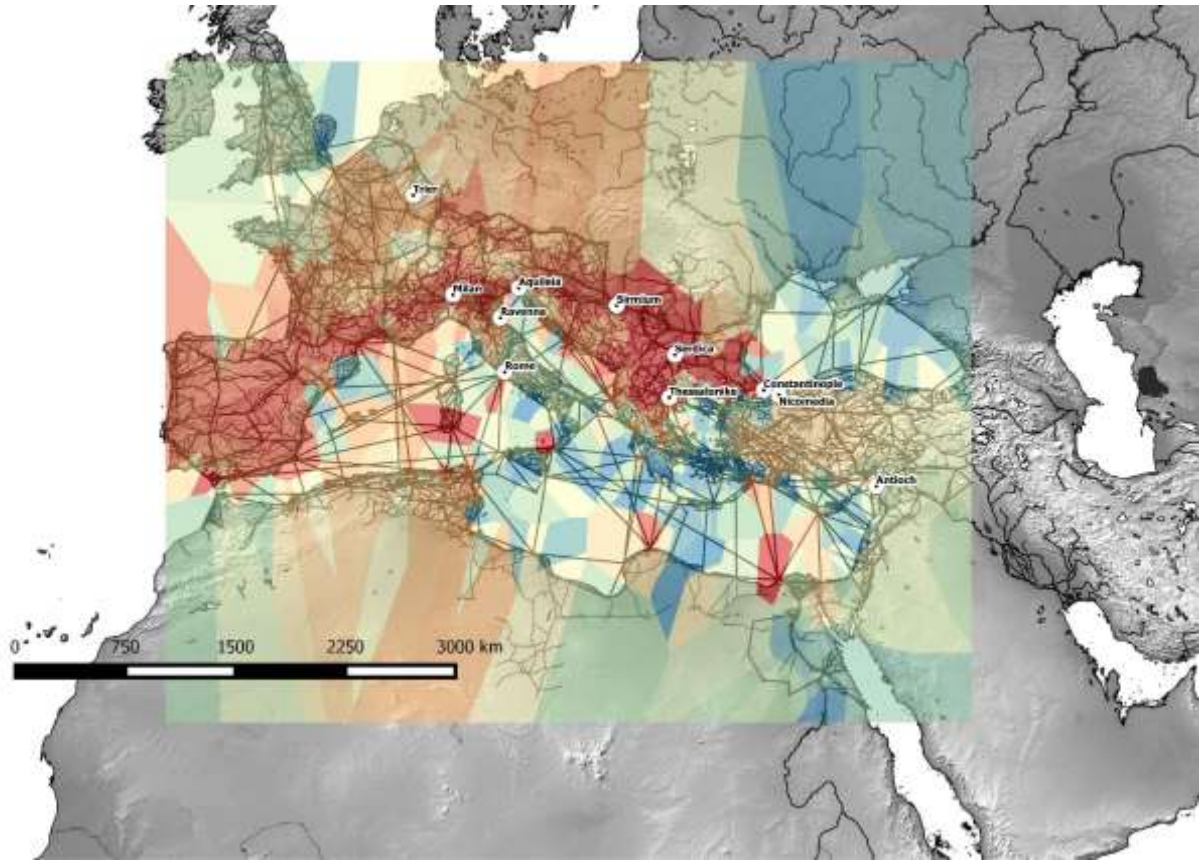

**Fig. 9:** Coloured Voronoi-map of the spatial distribution of closeness values of nodes in the ORBIS-network model for the Roman Empire (data: http://orbis.stanford.edu/; calculations and map: J. Preiser-Kapeller, 2018)

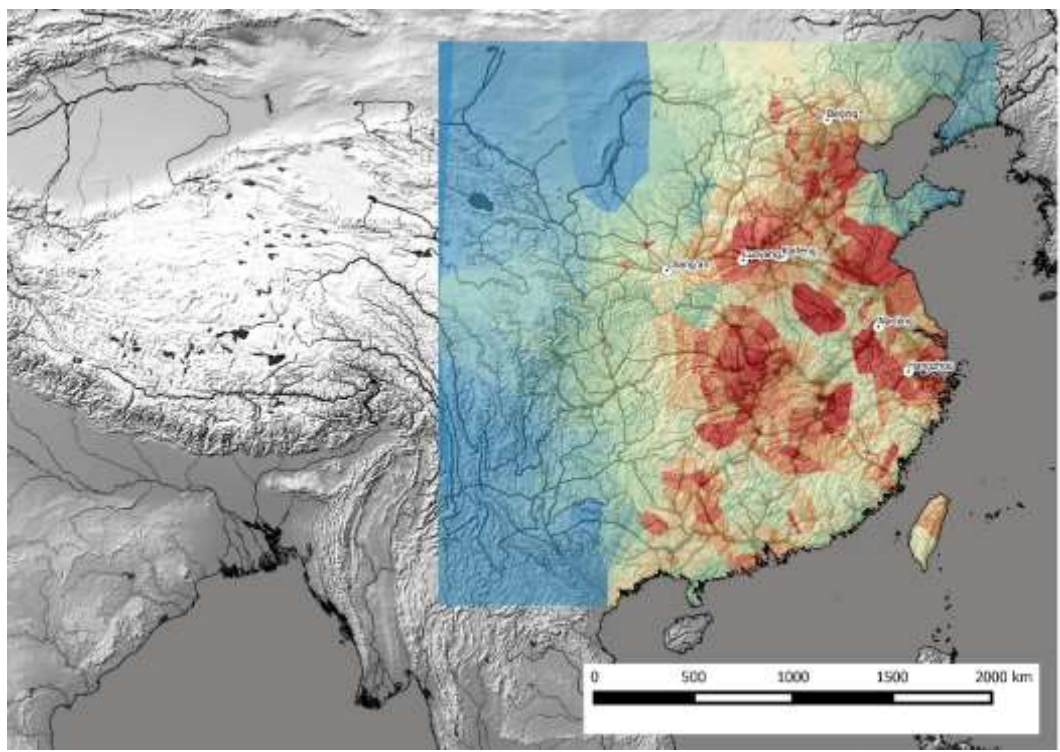

**Fig. 10:** Coloured Voronoi-map of the spatial distribution of closeness values of nodes in the network model for Imperial China (data: http://sites.fas.harvard.edu/~chgis/; calculations and map: J. Preiser-Kapeller, 2018)

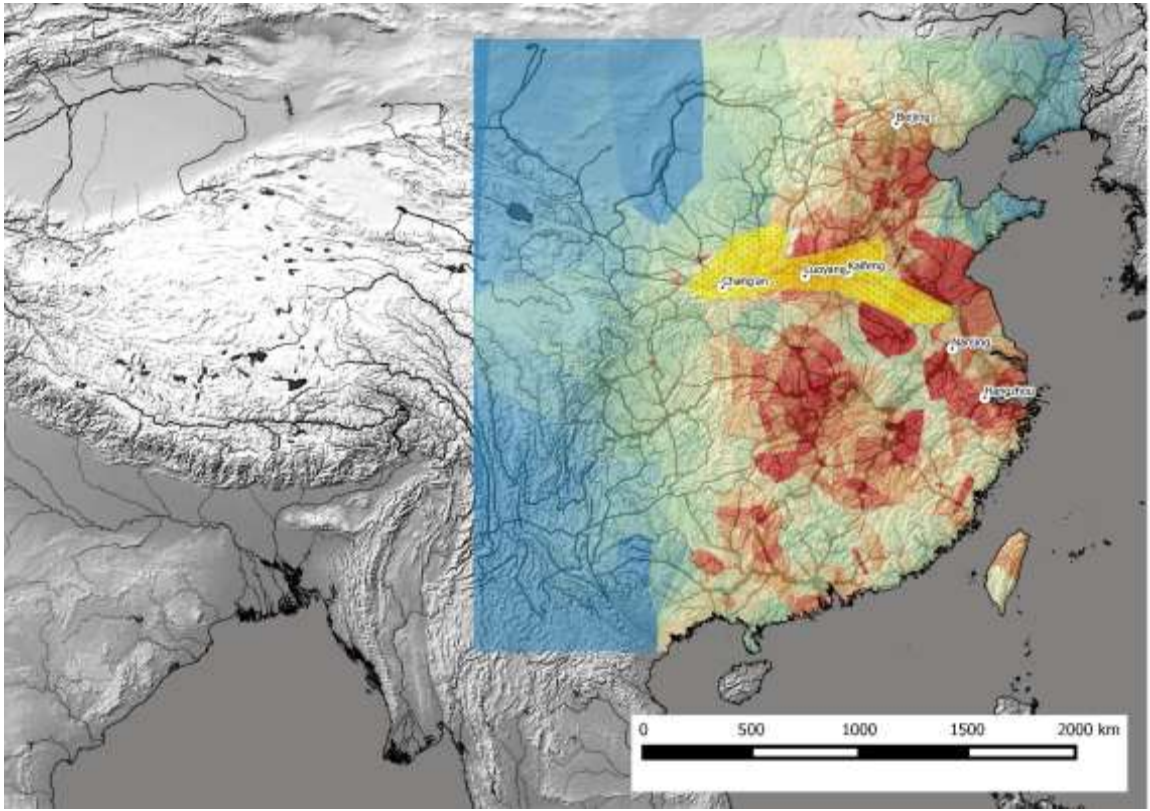

**Fig. 11:** The spread of a major epidemic in the years 636 to 643 CE (yellow) on a coloured Voronoi-map of the spatial distribution of closeness values of nodes in the network model for Imperial China (data: http://sites.fas.harvard.edu/~chgis/; calculations and map: J. Preiser-Kapeller, 2018)

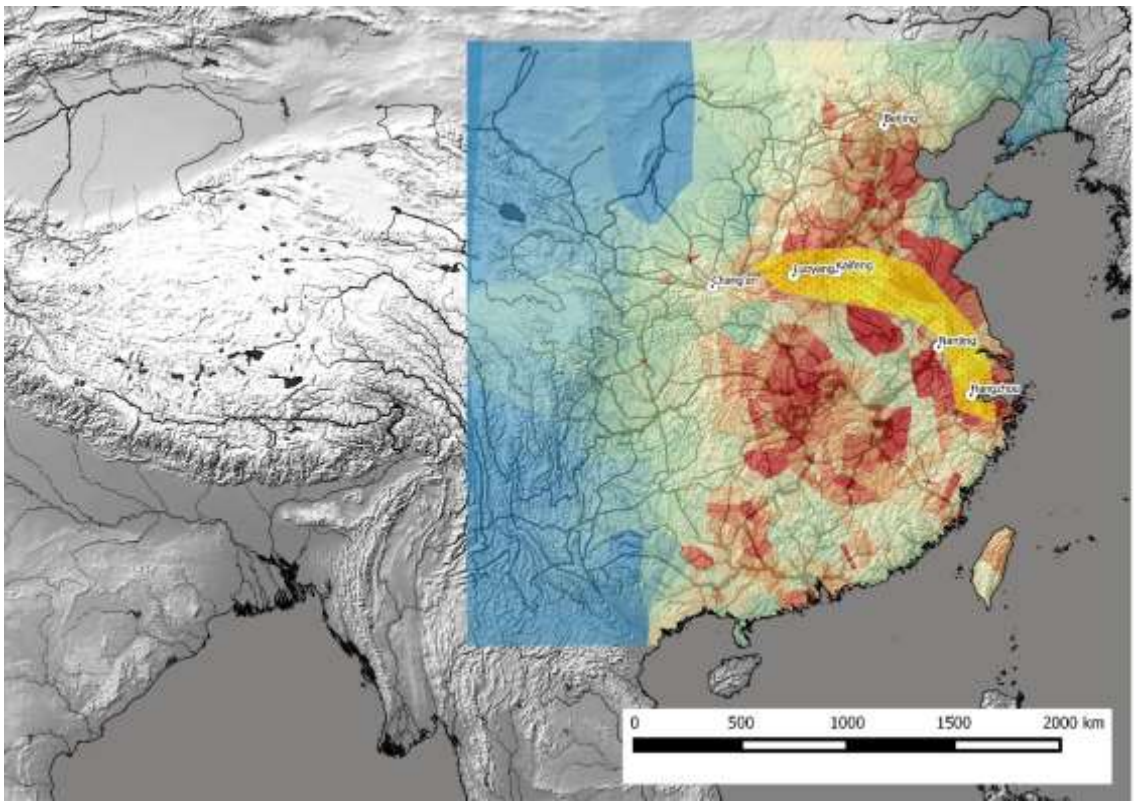

**Fig. 12:** The spread of a major epidemic in the year 1588 CE (yellow) on a coloured Voronoi-map of the spatial distribution of closeness values of nodes in the network model for Imperial China (data: http://sites.fas.harvard.edu/~chgis/; calculations and map: J. Preiser-Kapeller, 2018)

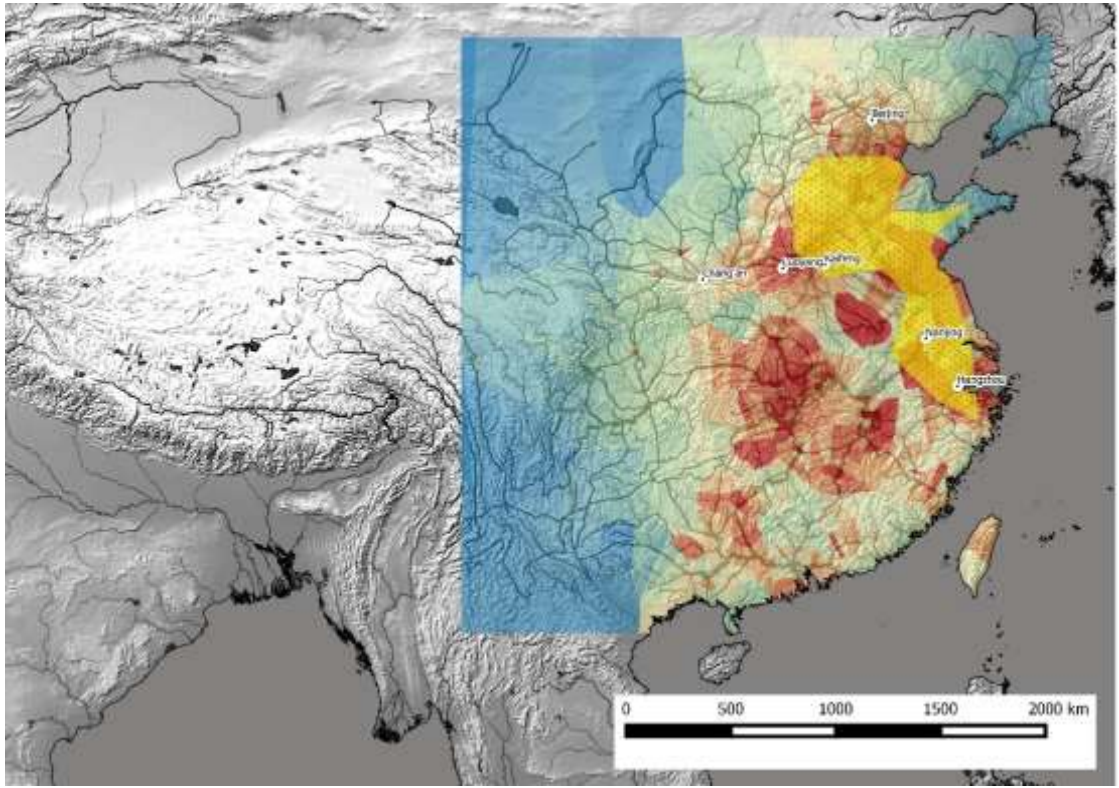

**Fig. 13:** The spread of a major epidemic in the year 1642 CE (yellow) on a coloured Voronoi-map of the spatial distribution of closeness values of nodes in the network model for Imperial China (data: http://sites.fas.harvard.edu/~chgis/; calculations and map: J. Preiser-Kapeller, 2018)

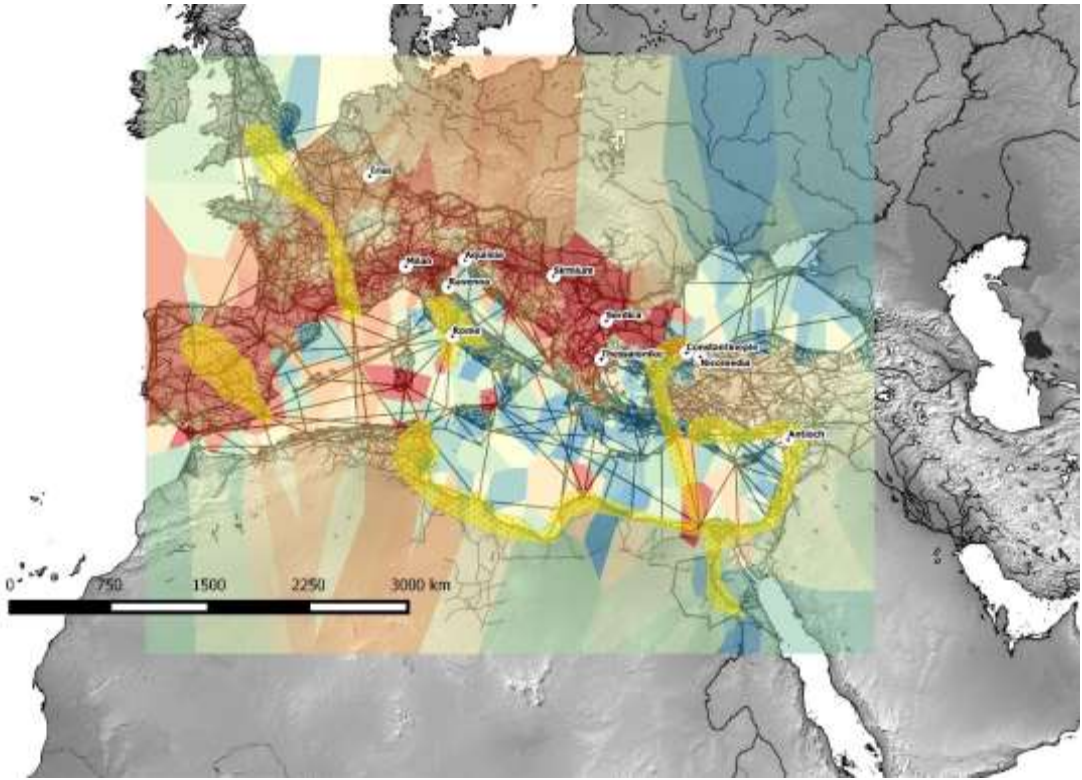

**Fig. 14:** Possible corridors of diffusion of Justinian´s plague in the 540s CE (yellow) on a coloured Voronoi-map of the spatial distribution of closeness values of nodes in the ORBIS-network model for the Roman Empire (data: http://orbis.stanford.edu/ and *Harper* 2017; calculations and map: J. Preiser-Kapeller, 2018)

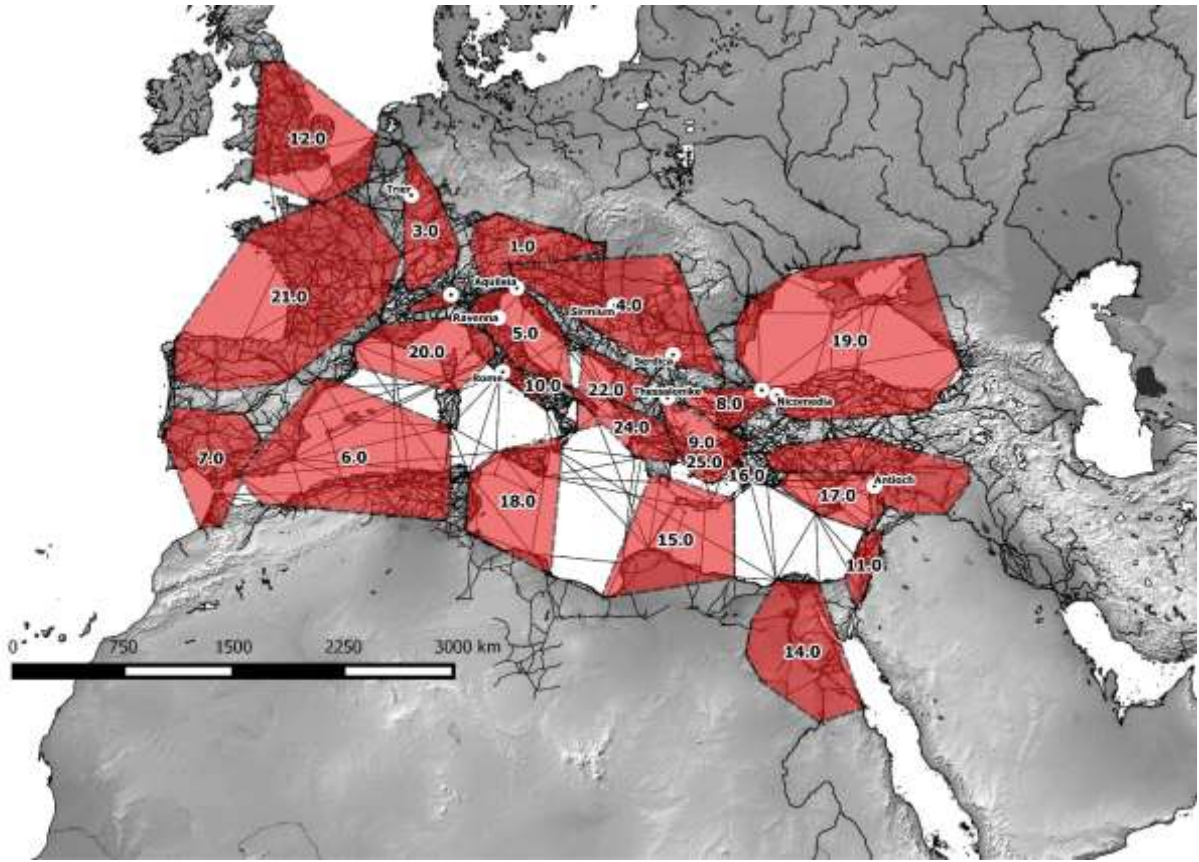

**Fig. 15:** 25 over-regional clusters identified among nodes in the ORBIS-network model for the Roman Empire with the help of the Newman-algorithm (data: http://orbis.stanford.edu/; calculations and map: J. Preiser-Kapeller, 2018)

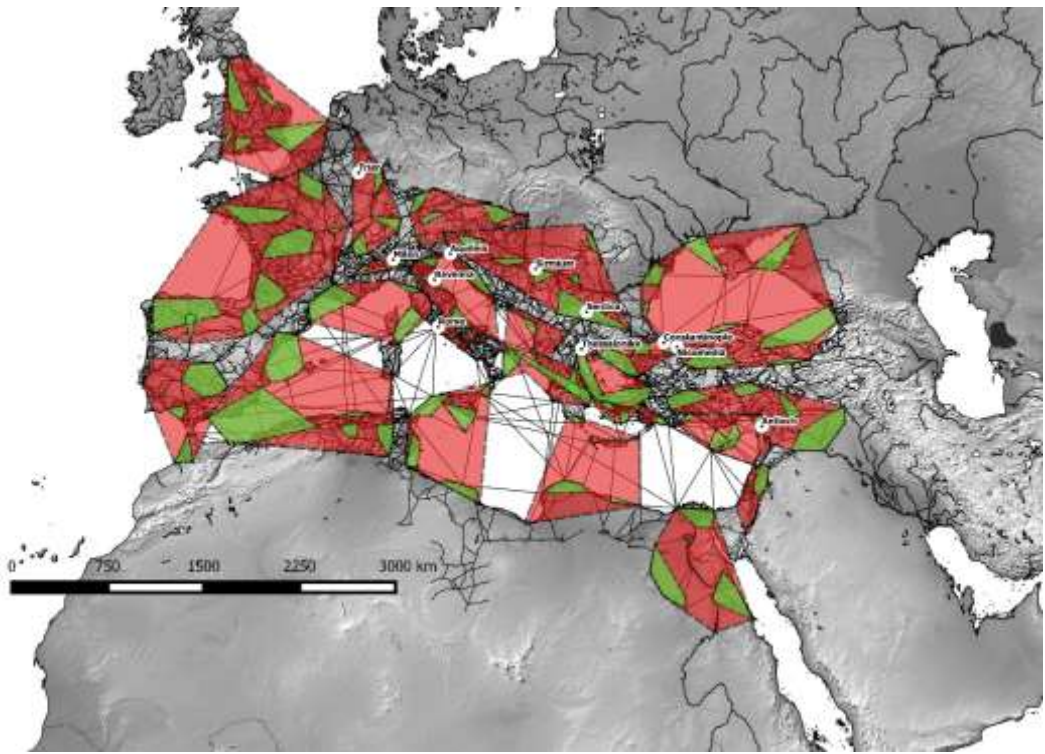

**Fig. 16:** Regional clusters (green) identified in the 25 over-regional clusters (red) identified among nodes in the ORBIS-network model for the Roman Empire with the help of the Newman-algorithm (data: http://orbis.stanford.edu/; calculations and map: J. Preiser-Kapeller, 2018)

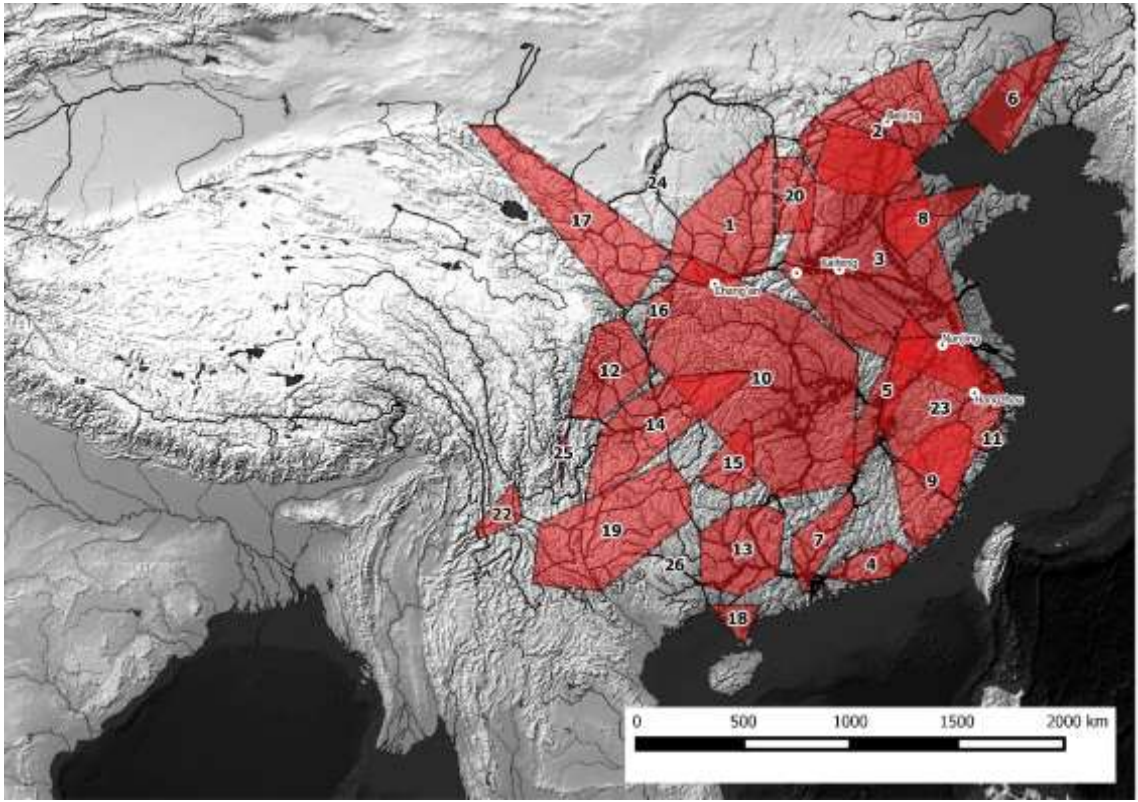

**Fig. 17:** 26 over-regional clusters identified among nodes in the network model for Imperial China with the help of the Newman-algorithm (data: http://sites.fas.harvard.edu/~chgis/; calculations and map: J. Preiser-Kapeller, 2018)

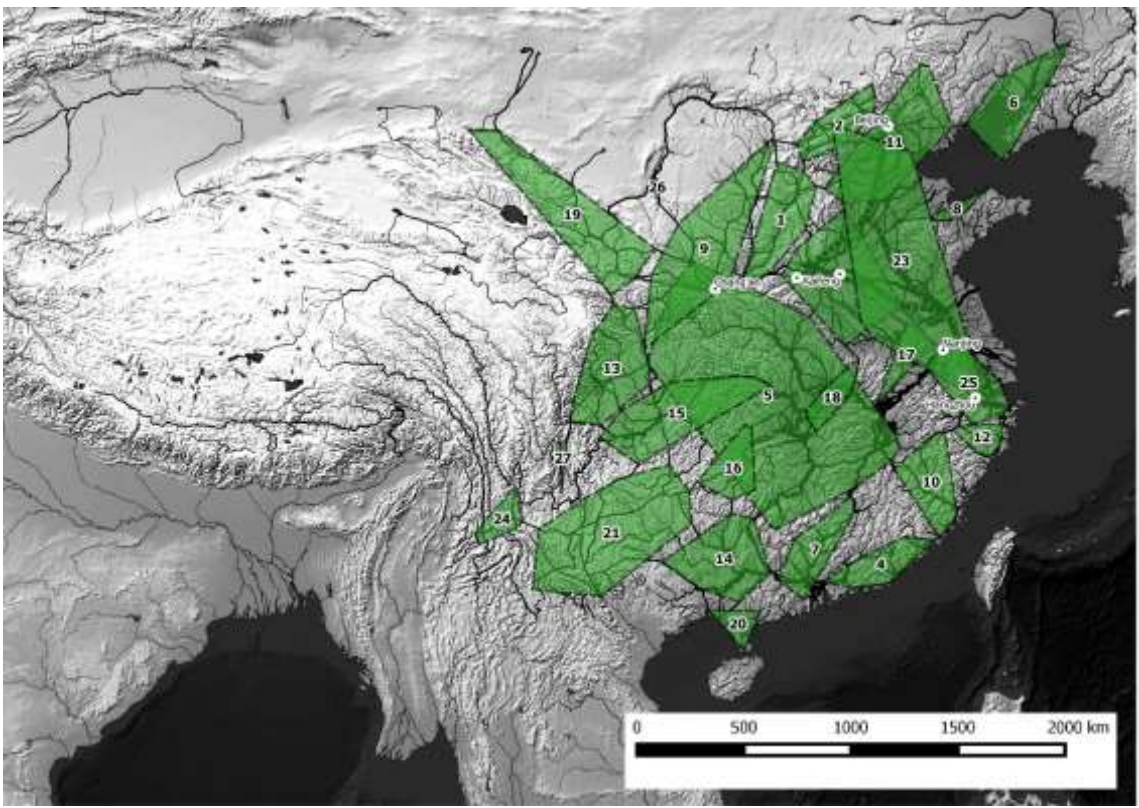

**Fig. 18:** 27 over-regional clusters identified among nodes in the network model for Imperial China with the help of the Louvain-algorithm (data: http://sites.fas.harvard.edu/~chgis/; calculations and map: J. Preiser-Kapeller, 2018)

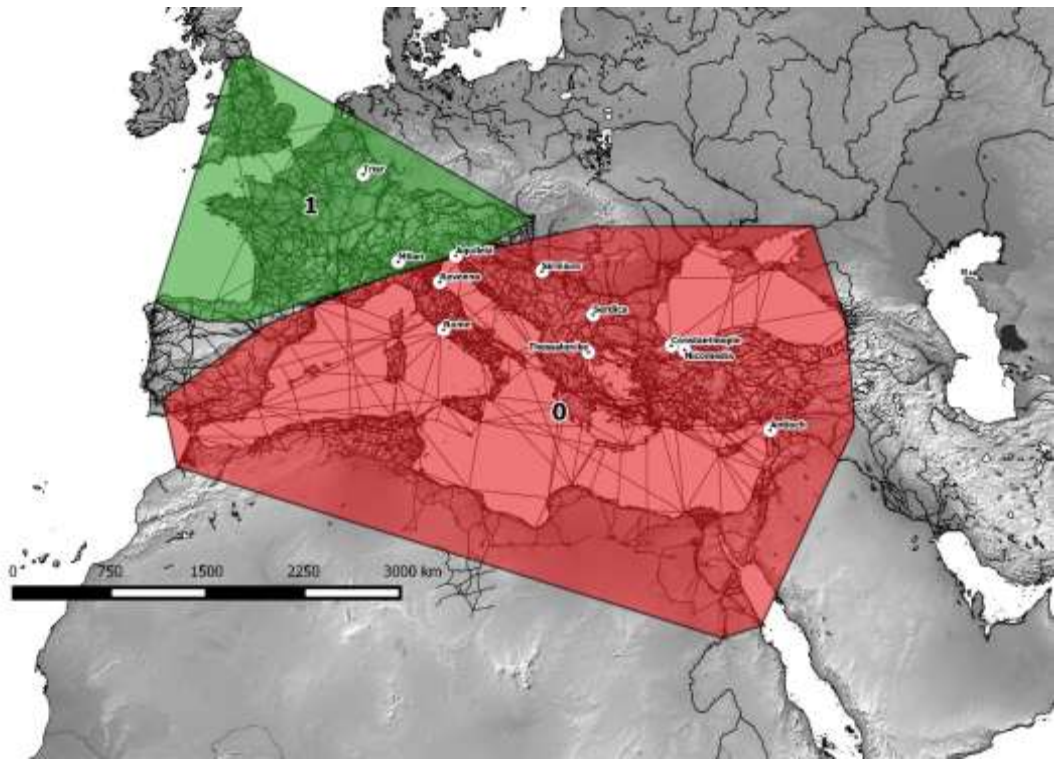

**Fig. 19:** Fragmentation of the ORBIS-network model for the Roman Empire in two components after the removal of the top 50 nodes in betweenness values (data: http://orbis.stanford.edu/; calculations and map: J. Preiser-Kapeller, 2018)

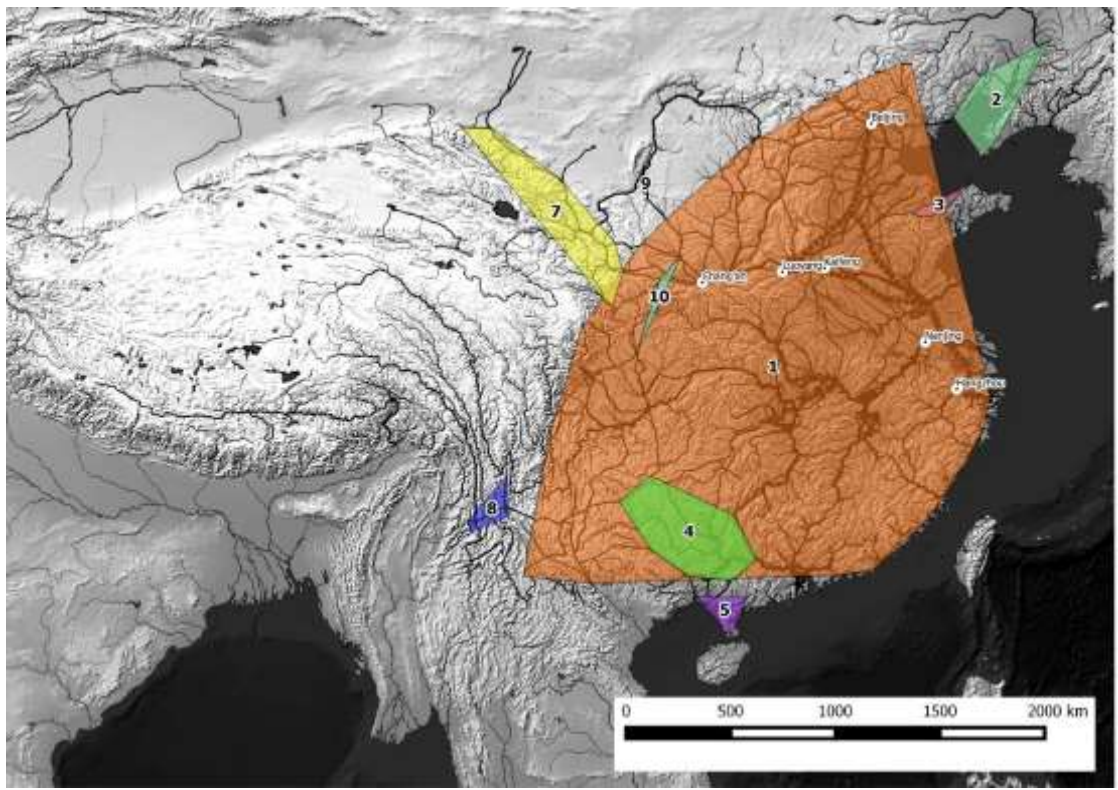

**Fig. 20:** Fragmentation of the network model for Imperial China in various components after the removal of the top 150 nodes in betweenness values (data: http://sites.fas.harvard.edu/~chgis/; calculations and map: J. Preiser-Kapeller, 2018)

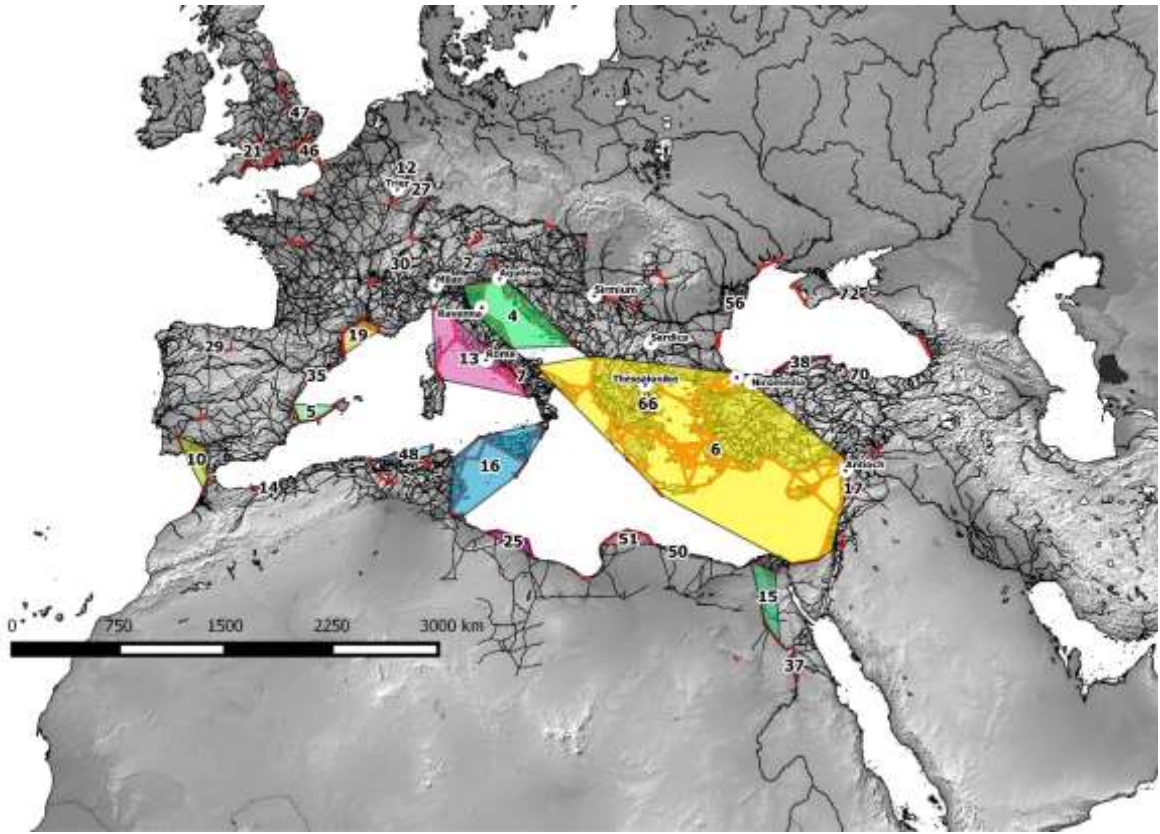

**Fig. 21:** Fragmentation of the ORBIS-network model for the Roman Empire in various components after the removal of all links beyond a cost-threshold of one day of travel (data: http://orbis.stanford.edu/; calculations and map: J. Preiser-Kapeller, 2018)

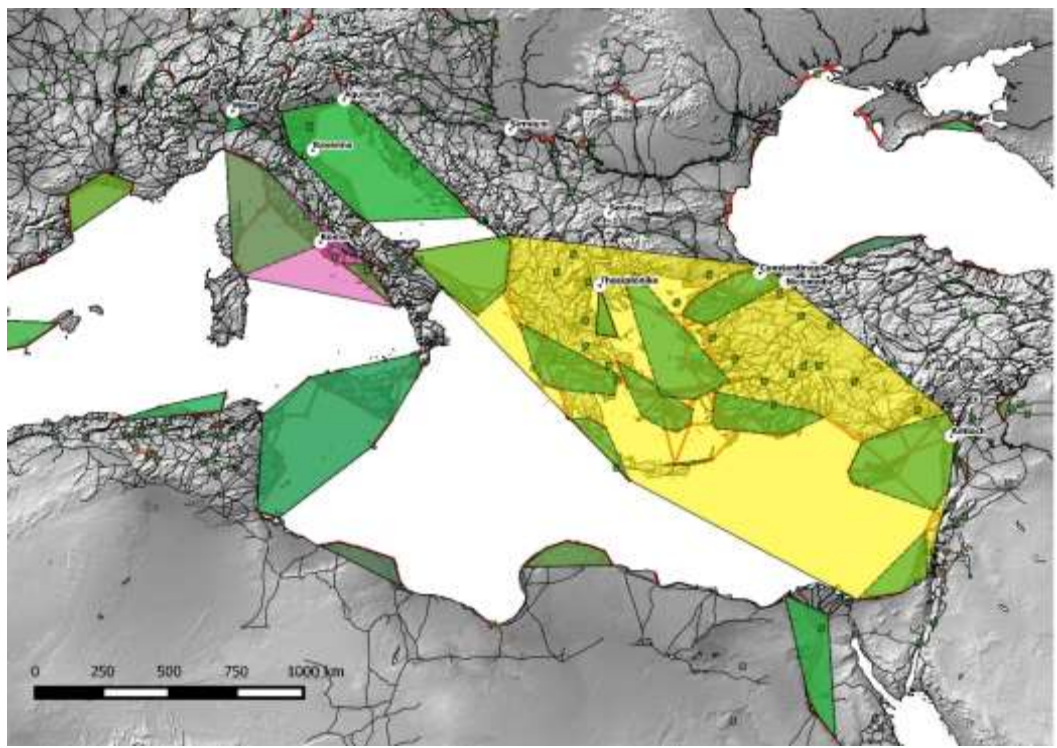

**Fig. 22:** Fragmentation of the ORBIS-network model for the Roman Empire in various components after the removal of all links beyond a cost-threshold of one day of travel and identification of regional clusters (in green) within these components with the help of the Newman-algorithm (data: http://orbis.stanford.edu/; calculations and map: J. Preiser-Kapeller, 2018)

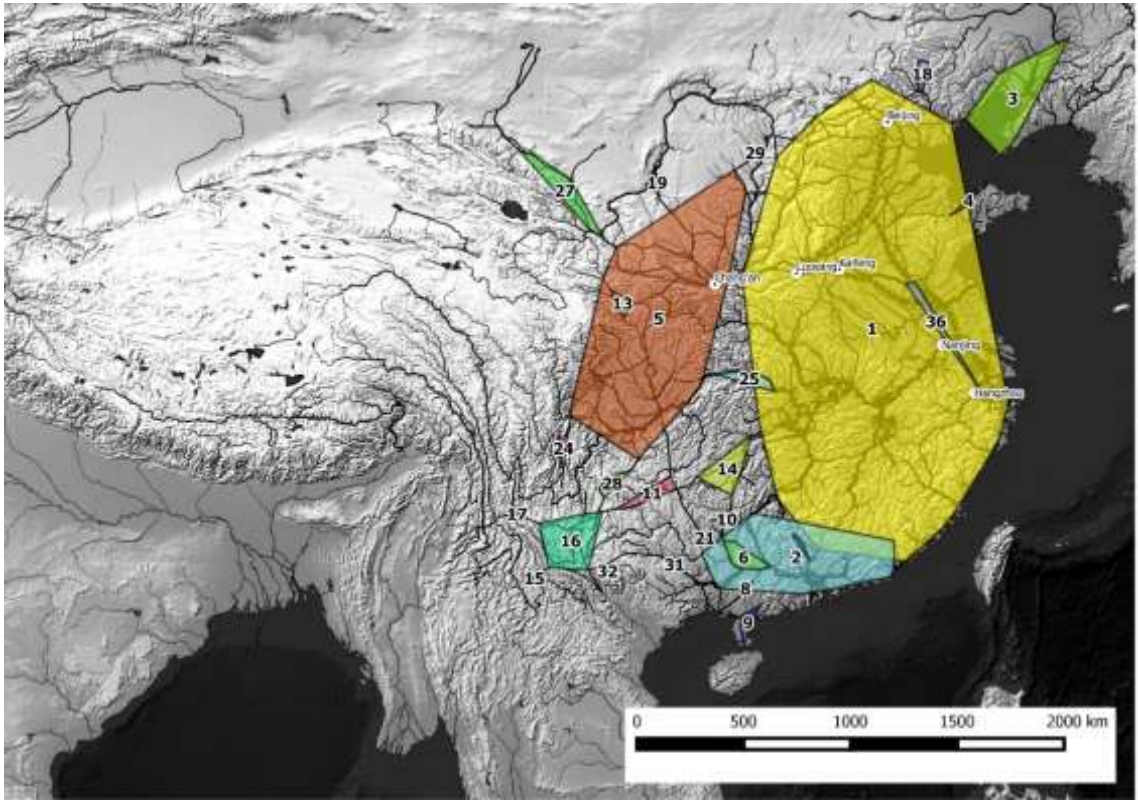

**Fig. 23:** Fragmentation of the network model for Imperial China in various components after the removal of all links beyond a cost-threshold of two days of travel (data: http://sites.fas.harvard.edu/~chgis/; calculations and map: J. Preiser-Kapeller, 2018)

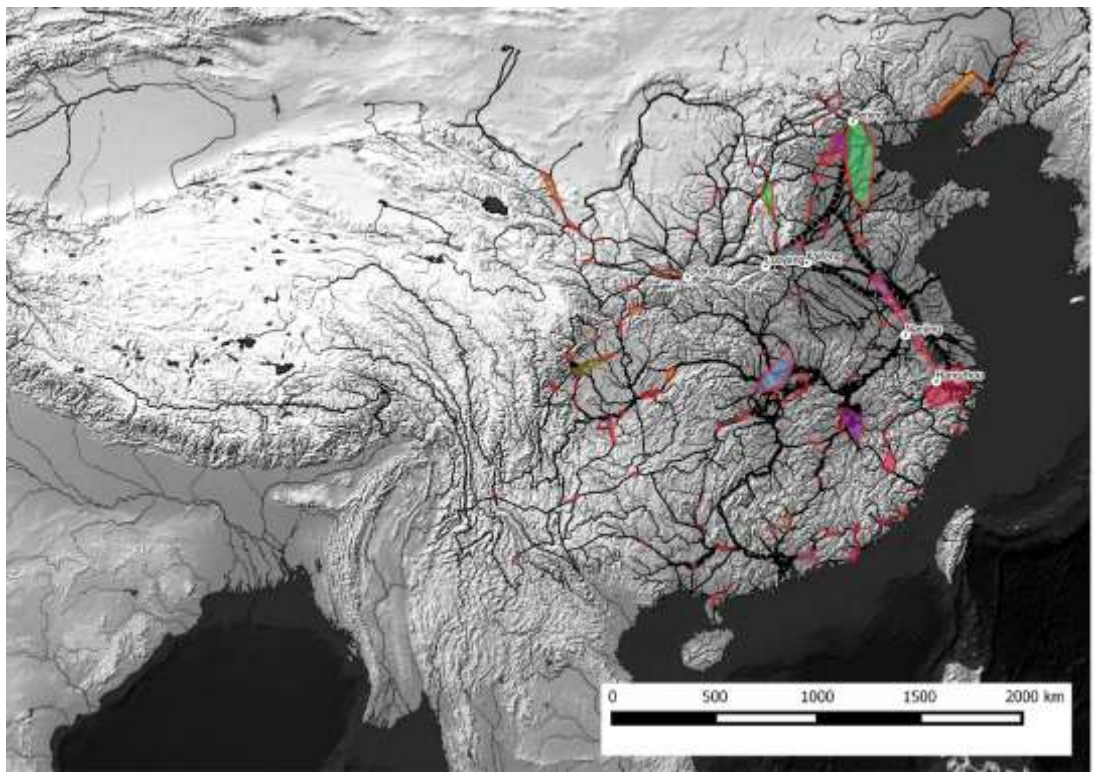

**Fig. 24:** Fragmentation of the network model for Imperial China in various components after the removal of all links beyond a cost-threshold of one day of travel (data: http://sites.fas.harvard.edu/~chgis/; calculations and map: J. Preiser-Kapeller, 2018)

41